\documentclass[useAMS,usenatbib,aastex]{mn2e}

\newcommand{\Sun}{\ensuremath{\odot}}
\newcommand{\e}{\ensuremath{\mathrm{e}}}

\newcommand{\N}{\ensuremath{\mathbb{N}}}

\usepackage[pdftex]{hyperref}
\usepackage{epsfig}
\usepackage{amsmath}
\usepackage{amssymb}
\usepackage{graphicx}
\usepackage{multirow}
\usepackage{array}
\usepackage{color}
\definecolor{gray}{gray}{0.7}

\title[Stellar orbits in cosmological simulations]
{Stellar orbits in cosmological galaxy simulations: the connection to formation history and line-of-sight kinematics}
\author[B.~R\"ottgers et al.]
{Bernhard R\"ottgers$^{1}$\footnotemark , Thorsten Naab$^{1}$ \& Ludwig Oser$^{2}$ \\
$^{1}$Max-Planck Institut f\"ur Astrophysik, Karl-Schwarzschild-Str. 1,
85741 Garching, Germany \\
$^{2}$Department of Astronomy and Astrophysics, Columbia University, New York, NY 10027, USA\\
}
\date{Accepted ???. Received ??? in original form ???}

\pagerange{\pageref{firstpage}--\pageref{lastpage}} \pubyear{2013}

\begin{document}
\label{firstpage}
\maketitle

\begin{abstract}
We analyze orbits of stars and dark matter out to three effective
radii for 42 galaxies formed in cosmological zoom simulations. Box
orbits always dominate at the centers and $z$-tubes become important at
larger radii.  We connect the orbital structure to 
the formation histories and specific features (e.g. disk,
counter-rotating core, minor axis rotation) in two-dimensional
kinematic maps. Globally, fast rotating galaxies with significant
recent {\it in situ} star formation are dominated by $z$-tubes. Slow rotators
with recent mergers have significant box orbit and $x$-tube
components. Rotation, quantified by the  $\lambda_R$-parameter often
originates from streaming motion of stars on $z$-tubes but sometimes
from figure rotation. The observed anti-correlation of $h_3$ and $V_0
/ \sigma$ in rotating galaxies can be connected to a dissipative
formation history leading to high $z$-tube fractions. For galaxies with recent mergers {\it in situ}
formed stars, accreted stars and dark matter particles populate
similar orbits. Dark matter particles have isotropic velocity
dispersions. Accreted stars are typically radially biased 
($\beta \approx 0.2 - 0.4$). {\it In situ} stars become tangentially biased
(as low as $\beta \approx -1.0$) if dissipation was
relevant during the late assembly of the galaxy. We discuss the
relevance of our analysis for integral field surveys and for
constraining galaxy formation models.
\end{abstract}


\footnotetext{E-mail: broett@mpa-garching.mpg.de}

\clearpage

\section{Introduction}
\label{intro}

The observable properties of the stellar components of early-type
galaxies (ETGs) result from the projected superposition of the light of
individual stars with given ages and metallicities on particular
orbits. Assuming that ETGs are in dynamical equilibrium the 
stars orbit in potentials which are mostly generated by the stars
themselves but also by dark matter (at larger radii) and in some cases
by a additional gaseous components. Theoretically, a large number of
equilibrium configurations can be constructed (see
e.g. \citealp{Binney_Tremaine_2nd}) but only a subset of all possible
models seem to be realized in nature. Important  information about the
relevant formation and assembly processes of ETGs might therefore be
stored in the orbital composition and the observable projected stellar
line-of-sight kinematics.            

Unlike for spiral galaxies, not all stars in present day 
massive ETGs have formed in the galaxy itself. It is more likely that
the early formation ($z \gtrsim 2$) is dominated by {\it in situ} star formation 
and the late assembly becomes dominated by stellar accretion
(see e.g. \citealp{2006ApJ...648L..21K,2008MNRAS.384....2G,2009ApJ...699L.178N,Oser_etal_2010, 
2011ApJ...736...88F,2012MNRAS.425..641L,2013MNRAS.428.3121M,2013MNRAS.429.2924H}). As
dissipation is important during the early phase the systems 
are very likely axisymmetric and it is expected that stars will mostly
form and move on tube orbits. Later on more and more stars (that have
formed in other galaxies) are accreted in major and minor mergers and
the final assembly of massive ETGs might be governed by collisionless
dynamics alone. Mergers, however, can significantly change the orbital
composition (and dynamics) of galaxies and direct information
about the early assembly process might be hidden or lost (see
e.g. \citealp{2005MNRAS.360.1185J}).  

But also long after the main epoch of star formation gas can have a
significant and observationally testable impact on the stellar
morphology and kinematics. For example, it has been demonstrated by
\citet{1996ApJ...471..115B} that during galaxy mergers gas can be
driven to the central regions of the merger remnants making the
central potential more axisymmetric. Such a configuration results in a
more axisymmetric stellar shape, disfavors the population of stars on
box orbits (which are the dominant orbit class for collisionless mergers
(e.g. \citealp{2005MNRAS.360.1185J})), and supports the population of
minor axis tube orbits
\citep{1996ApJ...471..115B,2006MNRAS.372..839N}. This change in
orbital configuration has a significant---and observable---impact on
the line-of-sight kinematics of galaxies. \citet{2006MNRAS.372..839N} 
showed that the presence of gas and the resulting change in orbital
configuration  changes the asymmetry of line-of-sight velocity
profiles. Collisionless remnants show steep trailing (retrograde)
wings (positive $h_3$) \citep{2001ApJ...555L..91N} whereas gas rich
remnants have steep leading (prograde) wings (negative $h_3$) in much
better agreement with observations
\citep{2006MNRAS.372..839N,2009ApJ...705..920H,2010ApJ...723..818H}.

Direct numerical cosmological simulations have become a valuable tool
to understand galaxy formation and are a major driver for theoretical
scientific progress in the field.  It is now possible to follow the
evolution of individual galaxies over a Hubble time at high
resolution, starting to resolve the internal stellar structure of
galaxies  (see e.g. \citealp{2007ApJ...658..710N}) and study the
impact of feedback processes on galaxy dynamics. For example, it has
been demonstrated that simulations with identical initial conditions,
but different feedback models can yield galaxies with significantly 
different morphological and kinematic properties
(e.g. \citealp{Piontek_Steinmetz_2011,   2012MNRAS.423.1726S,
  Puchwein_Springel_2013, 2013MNRAS.436.2929H,  Aumer_etal_2013}). Any  
form of `feedback' directly affects the distribution and thermodynamic
properties of the dissipative component (gas) in and around galaxies and thereby  
regulates when and where stars are formed---determining
their initial orbits---and how much gas is funneled to the centers or
expelled from the galaxies at which stage of their evolution. For the
same initial conditions, stronger feedback in general leads to more
{\it in situ} star formation, higher gas fractions, and less stellar
accretion at late times
\citep{2012MNRAS.419.3200H,2012MNRAS.425..641L,2013MNRAS.436.2929H}. This
will also lead to a changes in the orbital composition at the present
day.   

In this paper we provide the framework for a fast and self-consistent
analysis of stellar orbits in high-resolution cosmological zoom
simulations similar to \citet{2005MNRAS.360.1185J}. We focus on a set of
cosmological simulations with weak stellar feedback and test for
correlations with global and detailed  observable properties. The
impact of feedback on our results will be presented in a follow-up
study. In a
larger scale cosmological context, \cite{Bryan_etal_2012} have
presented a similar study on the orbital content of dark matter halos
of simulations with different  feedback and radiative cooling models
and indeed found the orbital content of the halo to change strongly with
different feedback models. 

In addition to the pure identification of the stellar orbits we
connect this information with observable properties showing up in
two-dimensional kinematic maps (see also \citet{Naab_etal_2013}). This
information provides direct theoretical input for the results from
existing and upcoming integral-field surveys like e.g. SAURON
\citep{2001MNRAS.326...23B},  ATLAS$^\text{3D}$
\citep{2011MNRAS.413..813C}, KMOS \citep{2006NewAR..50..370S}, VIRUS-P
\citep{2008SPIE.7014E.231H}, CALIFA \citep{2012A&A...538A...8S}, SAMI
\citep{2012ApJ...761..169F}, SLUGGS \citep{2014ApJ...791...80A}, MASSIVE 
\citep{2014arXiv1407.1054M} or MANGA.  Our results will help to
understand how the assembly history of a galaxy influences its orbital
content and the observable line-of-sight velocity distributions (LOSVDs) which are extracted from the survey
data.    

The paper is organized as follows: We briefly review the simulations
in Section~\ref{simulations} and present our analysis procedure
including the reconstruction of the potential, the orbit classification
and the construction of LOSV maps in Section~\ref{methods}. In 
Section~\ref{results} we present results on the connection of orbit 
classes with triaxiality, line-of-sight kinematics, LOSVD asymmetries, 
and the two-phase assembly process. We summarize and discuss our 
results in Section~\ref{summary}.

\section{Simulations}
\label{simulations}
In this paper we investigate the orbital structure of central
galaxies of the cosmological hydrodynamic zoom simulations
presented in \cite{Oser_etal_2010}. This sample of simulations 
(or parts of it) has been used to
study cosmological size and dispersion evolution
\citep{2012ApJ...744...63O}, the mass distribution
\citep{Lyskova_etal_2012} as well as the detailed  two-dimensional
kinematic properties at small \citep{Naab_etal_2013} and large radii
\citep{2014MNRAS.438.2701W}. The galaxies have present-day stellar 
masses
ranging from $6.1 \times 10^{10}\,M_\Sun$ to 
$6.9 \times 10^{11}\,M_\Sun$.

The simulations were performed using GADGET \citep{Springel_etal_2001, Springel_2005} with cosmological 
parameters based on the
three year results of the Wilkinson Microwave Anisotropy Probe (WMAP)
\citep{Spergel_etal_2007}: $\sigma_8 = 0.77$, $\Omega_m = 0.26$,
$\Omega_\Lambda = 0.74$, $h = 0.72$ and $n_s = 0.95$. The halos of
interest were drawn from a large-scale dark matter run with $512^3$
($\approx 10^8$)  particles in a cube with co-moving side length of
\mbox{$72$ Mpc $h^{-1}$} from redshift $z \simeq 43$ to $z = 0$ with a fixed
gravitational smoothing length of $2.52$ kpc $h^{-1}$ and a mass of
$m_p = 2 \times 10^8 M_\Sun$ per particle (see 
\cite{Moster_etal_2010}).

All particles of the halos of interest within $2 \times R_{200}$ at $z
= 0$ were traced back to the beginning of the simulation. A region 
enclosing all these particles at $z \simeq 43$ then was populated with
dark matter and gas particles ($\Omega_\text{dm} = 0.216$ and $\Omega_\text{b} = 0.044$) with smoothing lengths reduced to
\mbox{$0.89$ kpc $h^{-1}$} and \mbox{$0.4$ kpc $h^{-1}$},
respectively. The particle masses were $m_{*, \text{gas}} = 4.2
\times 10^6 M_\Sun$ for star and gas particles and $m_\text{dm}
= 2.5 \times 10^7 M_\Sun$ for dark matter particles. Outside this
region and beyond a `safety margin' of $1\,\text{Mpc}\,{h}^{-1}$ the
original dark matter particles were down-sampled depending on the
distance with sufficient resolution to represent long range forces. 

The simulations include the effect of a redshift-dependent uniform UV
background, radiative cooling, star formation, and energy feedback
from type-II supernovae using the model of \citet{Springel_Hernquist_2003}. 
As discussed in \citet{Oser_etal_2010,2012ApJ...744...63O} this
sub-grid model favors the efficient conversion of gas into stars at
high redshifts and supports the formation of spheroidal systems. 
However, the massive galaxies forming with this particular sub-grid
model have structural properties that agree reasonably well with
observed ellipticals (see also
\citealp{2007ApJ...658..710N, 2012ApJ...754..115J, Lyskova_etal_2012, Naab_etal_2013, 2014MNRAS.438.2701W}).
An overview of some global galaxy properties is given in Tab.~\ref{tab:gal_props} along with the global orbit fractions determined later in this work.

\section{Analysis}
\label{methods}

\subsection{Galaxy identification and orientation}

At $z=0$, we identify galaxies (including their dark matter halos)
with a friends-of-friends algorithm and determine their centers in
configuration space (position and velocity) using a shrinking sphere
technique  \citep{Power_etal_2003}. The central galaxies are then
defined by all baryonic particles within $R_{10}$, i.e.\ $10\%$ of the
virial radius $R_\text{vir} \equiv R_{200}$. 

\begin{figure}
\centering
\includegraphics[width = 0.23\textwidth]{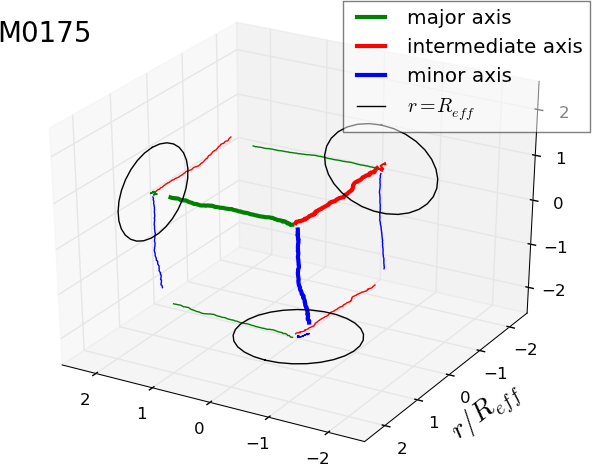} \,
\includegraphics[width = 0.23\textwidth]{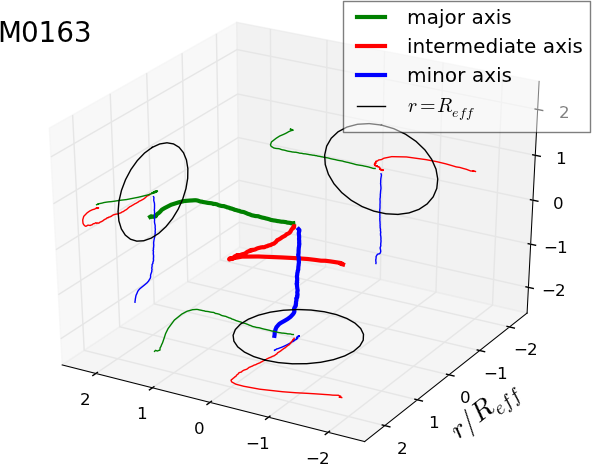} \,
\includegraphics[width = 0.23\textwidth]{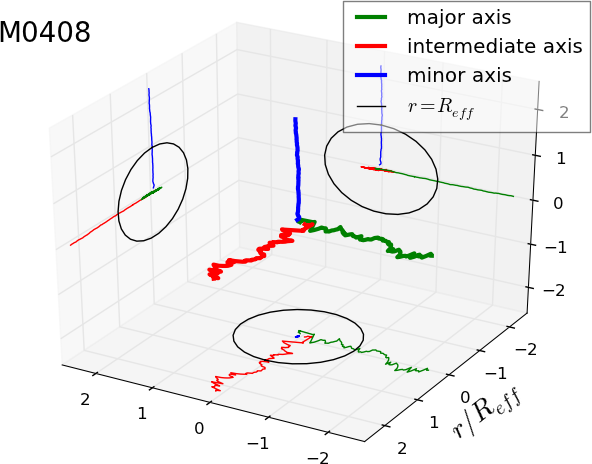} \,
\includegraphics[width = 0.23\textwidth]{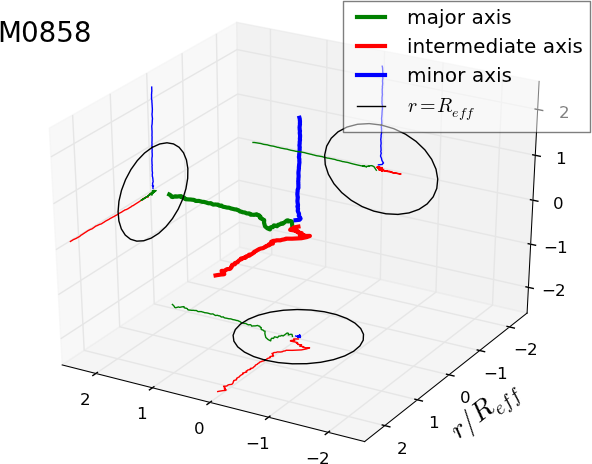}
\caption{The orientation of the eigenvectors of the reduced inertia tensor as a function of radius.
The black circles on the projections indicates the effective radius.
Most galaxies like M0175 and M0163 show good internal alignment and the orientation of the reduced inertia tensor only changes at larger radii, where large substructures are (cf.\ M0163).
Some very axisymmetric galaxies like M0408 have radially fluctuating eigenvectors in the plane perpendicular to the symmetry axis.
Only very few simulated galaxies show internal miss-alignment within $R_\text{eff}$, like M0858.}
\label{fig:orientation_stability}
\end{figure}

The orbit classification scheme (see Section~\ref{orbit_class})
requires the galaxies---or more precisely their potentials---to be 
oriented along their principal axes. Only a few degrees of 
misalignment can change the exact orbit fractions. We use the method 
presented in \citet{Bailin_Steinmetz_2005} to compute the `reduced 
inertia tensor' $\mathbf{\tilde I}$,  
\begin{align}
\tilde I_{ij} = \sum_{\text{Particles} \, k} m_k \, \frac{r_{k,i} \, r_{k,j}}{r_k^2},
\end{align}
where $m_k$ are the masses and $\vec r_k$ are the positions of the
individual stellar particles within a given radius; in this work within the 
half mass radius of the stars. The eigenvectors of this tensor then give 
the orientation of the principal axes.

An advantage of this method over the traditional moment-of-inertia
tensor is that it weighs particles at different distances from the center 
equally, whereas the normal moment-of-inertia tensor,
$I_{ij} = \sum_k m_k (\delta_{ij} r_k^2 - r_{k,i} r_{k,j})$, weighs masses
$m_k$ with $\mathcal O(r_k^2)$. We, however, would like to determine 
the orientation of the potential, whose depth is proportional to the mass, 
and hence the reduced inertia tensor is more appropriate. For example 
massive substructures at large radii would dominate the normal inertia 
tensor due to the $\mathcal O(r^2)$-factor, whereas the potential would 
still be dominated by the galaxy at small radii.

For most galaxies the orientation of the principal axes defined via
$\mathbf{\tilde I}$ does not vary significantly within the effective
radius\footnote{In this work we assume a constant light-to-mass ratio 
and we ignore gas. Hence, the effective radius is simply the 
half-mass radius of the stars.} (see Fig.~\ref{fig:orientation_stability}). 
Only a few galaxies show internal twists or misaligned cores like M0858. 
Galaxies that are almost axisymmetric like M0408 have the principal 
axis of the reduced inertia tensor perpendicular to the symmetry axis 
fluctuating with the radius up to which $\mathbf{\tilde I}$ was 
determined. However, the potential is axisymmetric as well and thus it
does not compromise the classification.

Substructures, that are massive enough to significantly impact the potential (basically on-going/up-coming mergers) and the direction of the principal
axis---like for M0163\mbox{---,} typically are located beyond 
$R_\text{eff}$.

We also compute the triaxiality parameter,
\begin{align}
T = \frac{1 - \left(b/a\right)^{2}}{1 - \left(c/a\right)^{2}}.
\end{align}
Here $a$ is the major axis, $b$ the intermediate axis and $c$ the minor 
axis. The two axis ratios can be calculated from the square roots, 
$\tilde a > \tilde b > \tilde c$, of the eigenvalues of the reduced inertia 
tensor \citep{Bailin_Steinmetz_2005}:
\begin{align}
\left(\frac{c}{a}\right) \simeq \left(\frac{\tilde c}{\tilde a}\right)^{\sqrt{3}} \quad \text{and} \quad \left(\frac{b}{a}\right) \simeq \left(\frac{\tilde b}{\tilde a}\right)^{\sqrt{3}}.
\end{align}

\subsection{Reconstruction of the potential}
\label{pot_reconstruction}

To classify the orbits of the stellar particles we need to know their
trajectories---at high temporal resolution---in the potential of the
galaxy. To speed up the orbital classification we freeze the system
at redshift zero, use a self-consistent field method (SCF,
\citealp{Hernquist_Ostriker_1992}) to extract the potential, and
then integrate the individual orbits of the particles we would like
to classify (for a similar approach see
\citealp{2005MNRAS.360.1185J, 2006MNRAS.372..839N, 2009ApJ...705..920H, Bryan_etal_2012}).

\begin{figure}
\begin{center}
\includegraphics[width = 0.5\textwidth]{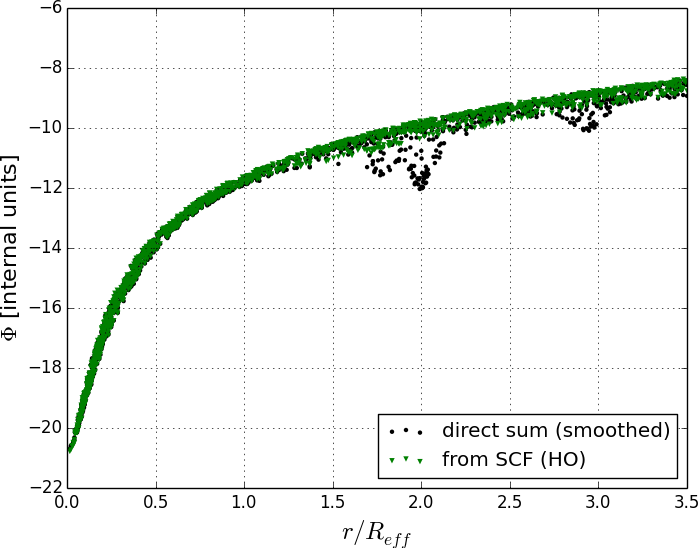}
\includegraphics[width = 0.5\textwidth]{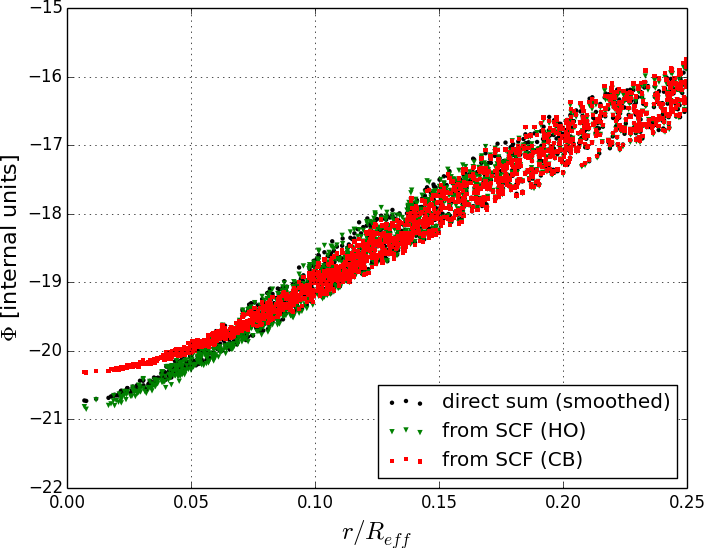}
\caption{{\it Top panel:} Comparison of the potential extracted from
  the particle distribution (black dots) and the SCF potential with a
  HO basis function set ($n_\text{max} = 18, \ell_\text{max} = 7$,
  green triangles) of galaxy M0163 for a set of randomly chosen
  stars. The fitted potential agrees well at the center and
  substructures at large radii are smoothed out. {\it Bottom panel:}
  Comparison of the direct (black) and HO (green triangles) potential to 
  a fit using the CB basis functions (red squares) using the same radial
  and angular order. At a fixed order the HO basis set fits the
  central potential much better. } 
\label{fig:scf_err}
\end{center}
\end{figure}

We have investigated two different basis functions for the potential
expansion. The first one (CB hereafter) was presented in
\citet{Clutton-Brock_1973}. At zeroth order it is the density and
potential of a Plummer sphere,  
\begin{align*}
&\rho(r) = \frac{3 M}{4 \pi a^3} \left( 1 + \frac{r^2}{a^2} \right)^{-5/2}, \\
&\phi(r) = - \frac{G M}{\sqrt{r^2 + a^2}},
\end{align*}
where $M$ is the total mass and $a$ is the scale parameter.
Another set of SCF basis functions (HO hereafter) was suggested by
\cite{Hernquist_Ostriker_1992}, for which the zeroth order density and 
potential are 
\begin{align*}
&\rho(r) = \frac{M}{2 \pi} \frac{a}{r} \frac{1}{\left( r + a \right)^3}, \\
&\phi(r) = - \frac{G M}{r + a}.
\end{align*}
This is the Hernquist density profile \citep{Hernquist_1990} which is
a popular model for representing spheroidal galaxies
(e.g. \citealp{2013MNRAS.429.2924H}) and resembles a de
Vaucouleur's $R^{1/4}$ surface density profile (see however
\citealp{2006MNRAS.369..625N}).  

For both basis functions the difference of the real (simulated)
potential and the zeroth order term is expanded in a bi-orthogonal
basis set built-up from the spherical harmonics and a radial basis
set. The better the first order terms follow the underlying model, the
less terms in the basis set expansion are needed to produce a good
approximation to the original potential. We have tested both basis set
for varying radial and angular expansion terms and in all cases we
find that the HO basis set results in a significantly better
representation of the potential as it does not have a flat density
core inside the scale radius. In Fig.~\ref{fig:scf_err} we give an
example of the potential reconstruction for galaxy M0163. The SCF
potential and the potential from the simulation agree very well. In
addition, substructures at larger radii are washed out which results in
a smoother overall potential and makes the orbit classification more 
stable. At the center, the HO basis set represents the potential much 
better than the CB basis set (bottom panel of Fig.~\ref{fig:scf_err}). 
Moreover the quality of the HO fit is less sensitive to the choice of the 
scale parameter $a$ and the typical relative error of this basis set
assuming a radial order $n_\text{max} = 18$ and an angular order
$\ell_{max} = 7$ is less than $0.5\%$. Hence, we use the HO basis
function set to reconstruct the potential of all simulated galaxies.

\subsection{Orbit classification} 
\label{orbit_class}

In integrable systems bound orbits are confined on manifolds of
phase space that are topologically equivalent to tori. This allows
one to describe them with angle-action variables $(\boldsymbol \theta,
\boldsymbol J)$ in which the equations of motion become
\begin{align}
0 = \dot J_i &= - \frac{\partial \mathcal H}{\partial \theta_i} \\
\dot \theta_i &= - \frac{\partial \mathcal H}{\partial J_i} =: \Omega_i(\boldsymbol J),
\end{align}
where $\mathcal H$ is the Hamiltonian. The $\Omega_i$ are called
fundamental frequencies and are conserved quantities. It can be shown
that real space trajectories $\boldsymbol x(t)$ are quasi-periodic,
i.e.\  they can be written in a Fourier series,
\begin{align}
\boldsymbol x(t)
  = \sum\limits_{\mathbf n \in \N^3} \mathbf X_{\mathbf n}(\boldsymbol J) \,
  \e^{\mathrm i \mathbf n \cdot \boldsymbol \theta(t)},
\end{align}
with constant amplitudes $\mathbf X_{\mathbf n}$ and linearly growing angles $\boldsymbol \theta(t) = \boldsymbol \theta(0) + \boldsymbol \Omega \cdot t$. Such quasi-periodic
orbits are also called regular orbits. 

Galaxies do not have to be integrable systems. However, orbit theory
and in particular KAM theory (see e.g.\ \citealp{Kolmogorov_1954, Kolmogorov_1954_en,Moser_1973, Arnold_1989, Binney_Tremaine_2nd, Arnold_etal_2007}) tell us, that
near-integrable systems (whose Hamiltonian differs only slightly from
an integrable Hamiltonian) still have `islands' of  regular orbits
next to their otherwise irregular/chaotic orbits. The smaller the
deviation of the Hamiltonian from an integrable one, the larger these
island are. 

\begin{figure*}
\centering
\begin{minipage}{0.12\textwidth}
irregular orbit \\
(from M0175)
\end{minipage}
\raisebox{-.5\height}{
\includegraphics[width = 0.21\textwidth]{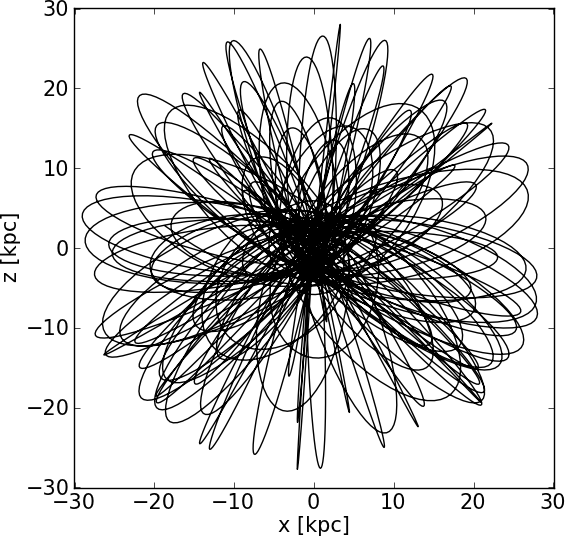} \ 
\includegraphics[width = 0.21\textwidth]{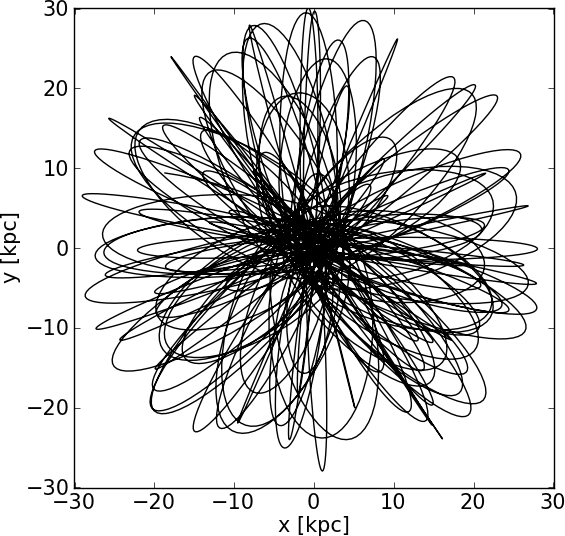} \ 
\includegraphics[width = 0.21\textwidth]{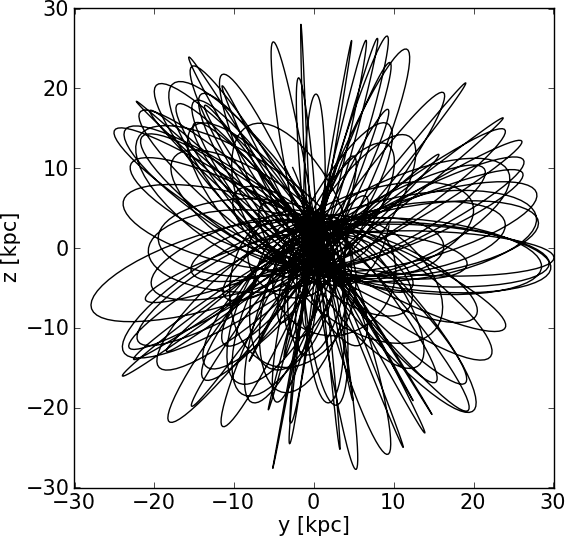}
}
\vspace{0.10cm}

\begin{minipage}{0.12\textwidth}
$\pi$-box orbit \\
(from M0175)
\end{minipage}
\raisebox{-.5\height}{
\includegraphics[width = 0.21\textwidth]{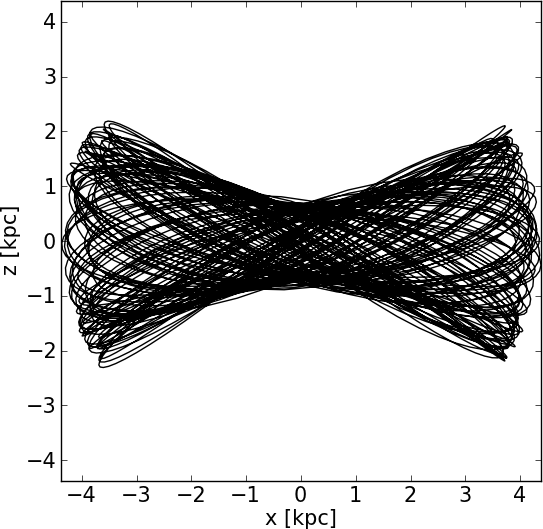} \ 
\includegraphics[width = 0.21\textwidth]{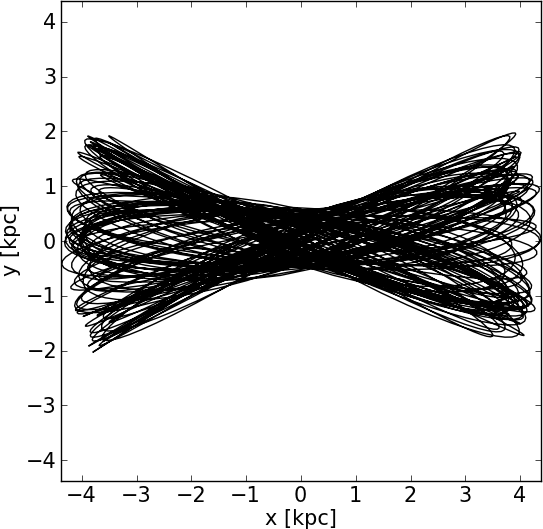} \ 
\includegraphics[width = 0.21\textwidth]{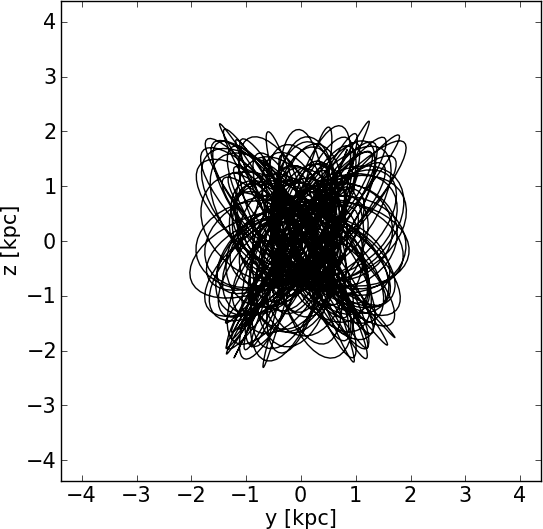}
}
\vspace{0.10cm}

\begin{minipage}{0.12\textwidth}
boxlet \\
(from M0190)
\end{minipage}
\raisebox{-.5\height}{
\includegraphics[width = 0.21\textwidth]{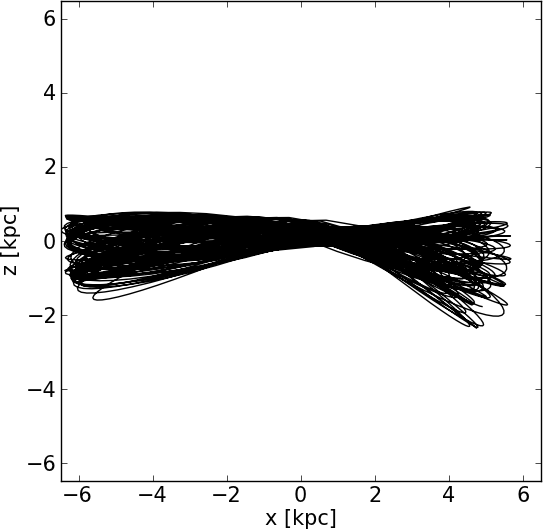} \ 
\includegraphics[width = 0.21\textwidth]{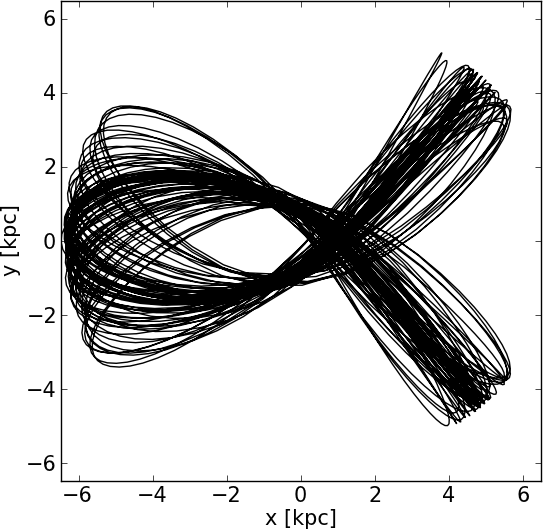} \ 
\includegraphics[width = 0.21\textwidth]{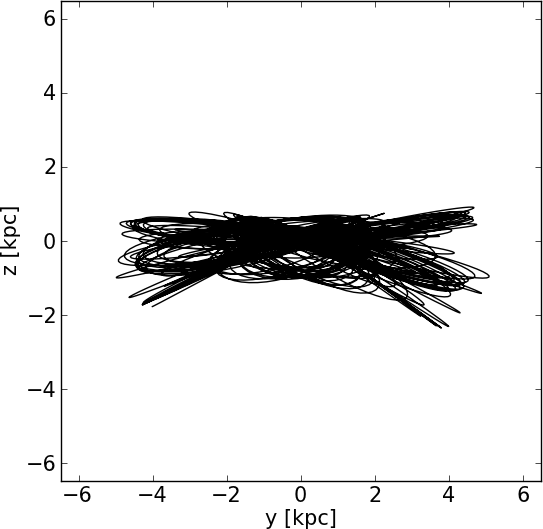}
}
\vspace{0.10cm}

\begin{minipage}{0.12\textwidth}
$z$-tube \\
(from M0175)
\end{minipage}
\raisebox{-.5\height}{
\includegraphics[width = 0.21\textwidth]{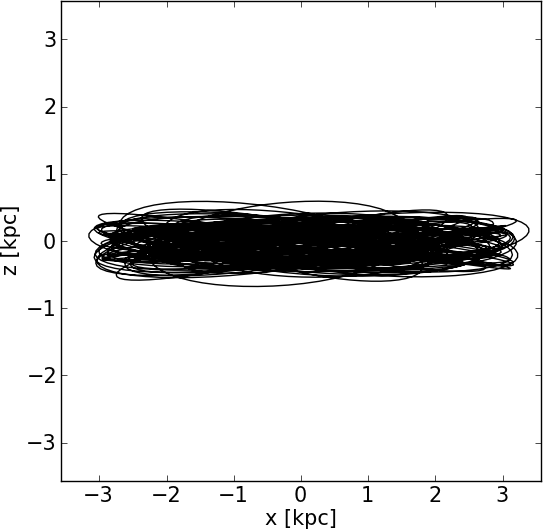} \ 
\includegraphics[width = 0.21\textwidth]{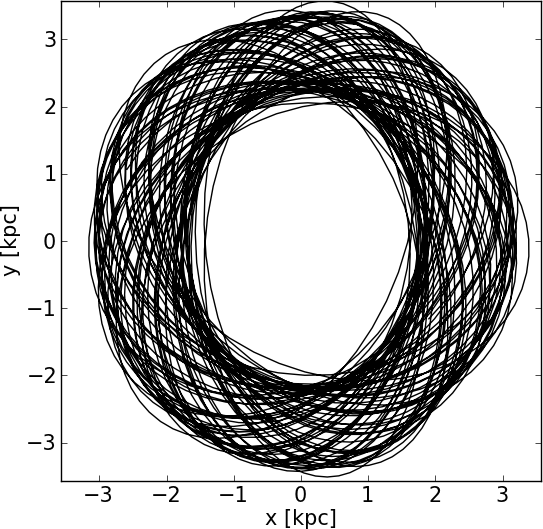} \ 
\includegraphics[width = 0.21\textwidth]{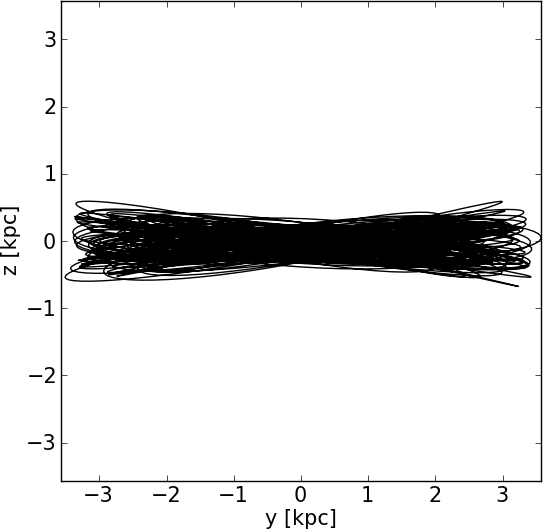}
}
\vspace{0.10cm}

\begin{minipage}{0.12\textwidth}
inner $x$-tube \\
(from M0125)
\end{minipage}
\raisebox{-.5\height}{
\includegraphics[width = 0.21\textwidth]{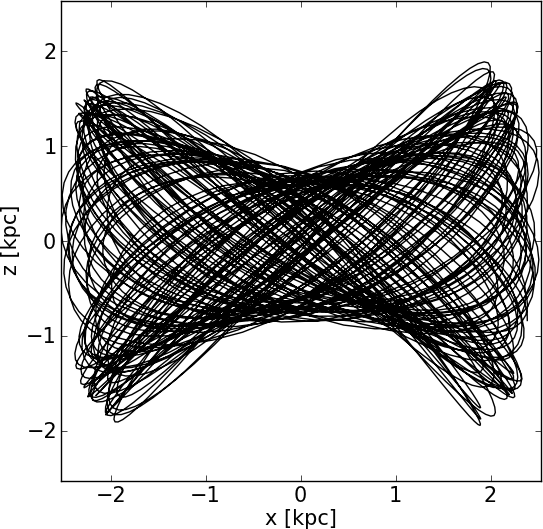} \ 
\includegraphics[width = 0.21\textwidth]{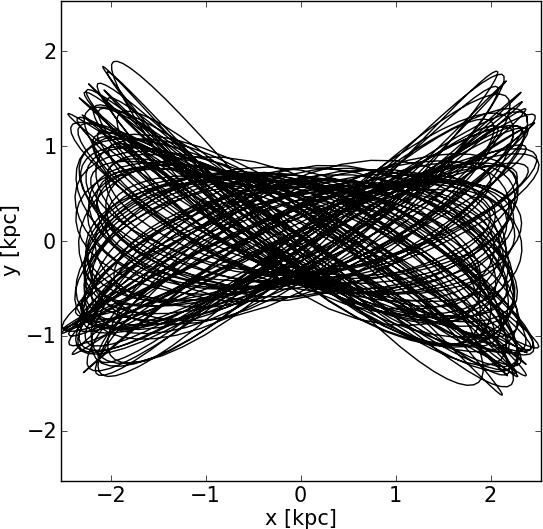} \ 
\includegraphics[width = 0.21\textwidth]{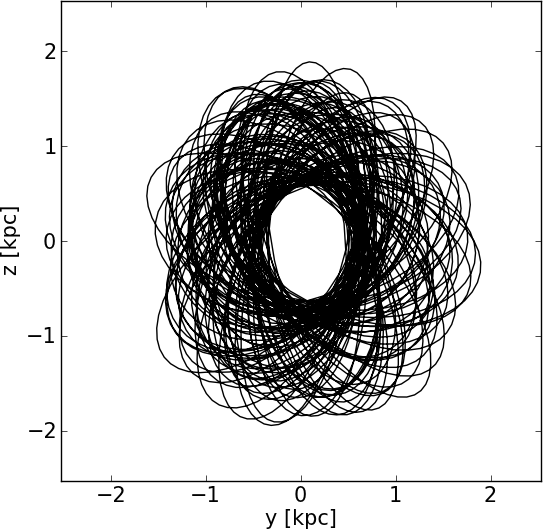}
}
\vspace{0.10cm}

\begin{minipage}{0.12\textwidth}
outer $x$-tube \\
(from M0175)
\end{minipage}
\raisebox{-.5\height}{
\includegraphics[width = 0.21\textwidth]{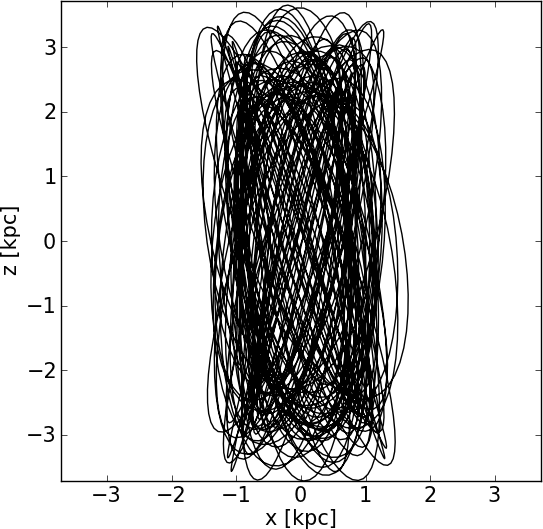} \ 
\includegraphics[width = 0.21\textwidth]{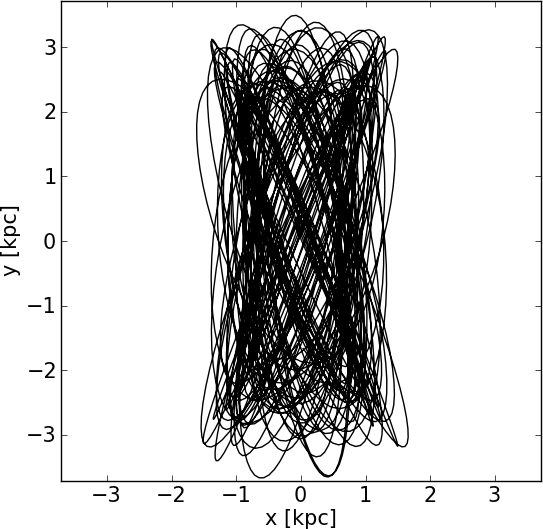} \ 
\includegraphics[width = 0.21\textwidth]{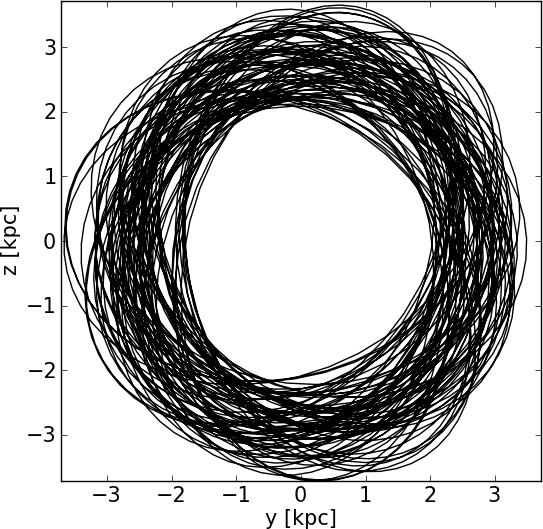}
}
\caption{Selected examples of the most regular trajectories (for 
  three projections along the principal axes, from left to right) of the 
  main orbit classes for individual particles taken from different 
  simulated galaxies complemented with one very chaotic/irregular 
  orbit. All tube orbits are centrophobic, while box orbits and irregular
  orbits can go through the center.}
\label{fig:classes_traj}
\end{figure*}

It turns out that regular orbits can be classified into four different
main classes, depending on the ratios of the $\Omega_i$ (see e.g.\ 
\citealp{de_Zeeuw_1985, Statler_1987, Binney_Tremaine_2nd}): box
orbits, minor-axis orbits (we call them $z$-tubes for short) and inner and 
outer major-axis orbits ($x$-tubes for short, see Fig.~\ref{fig:classes_traj}). 
These classes have fundamentally different physical properties. Box 
orbits have no mean angular momentum and are typically centrophilic, 
whereas all tube orbits are centrophobic and have a net angular 
momentum which is either  aligned with the $z$-axis (for $z$-tubes) or with 
the $x$-axis (for $x$-tubes). They also have different, often interesting 
geometries (e.g.\ the `boxlet' in Fig.~\ref{fig:classes_traj}). 

\cite{Carpintero_Aguilar_1998} (CA98 hereafter) developed a
classification scheme, that almost entirely builds on the detection of
resonances (only inner and outer $x$-tubes have to be distinguished
morphologically). $z$-tubes, for instance, have a 1:1 resonance in
the two dominant frequencies in the $x$- and the $y$-coordinate.
CA98 have also written a program to automatically classify orbits by
using the Fourier transforms of the trajectories (in three
dimensions), detecting the most prominent lines and then
checking  these lines for resonances. Depending on the number of base 
frequencies\footnote{Base frequencies are very similar to the
  fundamental frequencies, but not necessarily the same. It might be
  that one of them is twice the corresponding fundamental frequency,
  for instance, and the amplitude of uneven multiples of the
  fundamental frequency are undetectable small.}
that are needed to build up the detected lines (by their linear
combinations) and the number and kind of resonances the orbits 
are classified (see Tab.~\ref{tab:class_scheme}).

\begin{table}
\centering
\begin{tabular}{|c|c||c|c|c|}
\hline
\multicolumn{1}{|c}{} & \multicolumn{1}{c||}{} & \multicolumn{3}{c|}{number of base frequencies} \\
\cline{3-5}
\multicolumn{1}{|c}{} & & $<3$ & $=3$ & $3<$ \\
\hline
\hline
\multirow{2}{*}{number} & $0$ & low dim. & 3-D $\pi$-box & \multirow{5}{*}{irregular} \\
\cline{2-2}\cline{4-4}
 & \multirow{2}{*}{$1$} & (closed & open $\pi$:$m$:$n$ box & \\
of & & and & open $\pi$:$1$:$1$ tube & \\
\cline{2-2}\cline{4-4}
\multirow{2}{*}{resonances} & \multirow{2}{*}{$3$} & thin) & open $l$:$m$:$n$ box & \\
 & & orbits & open $l$:$1$:$1$ tube & \\
\hline
\end{tabular}
\caption{Summary of the spectral orbit classification in three
  dimensions as defined in
  \protect\cite{Carpintero_Aguilar_1998}. Chaotic orbits have more
  than three base frequencies; regular orbits are classified according
  to their resonance: tubes, boxlets ($\pi$:$m$:$n$ or $l$:$m$:$n$
  boxes) or (non-resonant) $\pi$-boxes. Depending on whether their
  trajectories fill a three dimensional volume (`open' orbits) or a
  lower dimensional one, they have three or less base frequencies,
  respectively.} 
\label{tab:class_scheme}
\end{table}

This scheme is capable of differentiating between $z$-tubes, $x$-tubes,
non-resonant box orbits---so called \mbox{`$\pi$-boxes'}---and resonant box orbits
(all resonance apart from 1:1 resonances)---called `boxlets'---, as well
as second order resonances. These occur when a particle is oscillating
resonantly around a stable, already resonant orbit. We, however, only 
make use of the differentiation between the main orbit classes (boxes,
$z$-tubes, $x$-tubes, irregulars) and non-classified orbits.  

The basic classification procedure works as follows: first we
integrate the particles, starting from their coordinates and
velocities at $z = 0$ in the reconstructed SCF potential with a
Runge-Kutta integrator of 8th order with adaptive time steps (capable
of continuous  on-the-fly output). We require a relative accuracy in
position and velocity of $10^{-8}$ per step, which leads to relative
changes in the particles' energies of order $10^{-5}$. The CA98
code then actually classifies an orbit three times
using slightly different parts of the trajectory. If all three
classifications differ the particle counts as `not classified'. 

The orbit classification for the almost $3 \times 10^6$ star particles in 
this work would have taken several months with the original serial
code. Therefore we have parallelized the loop over the particles with 
OpenMP and with that reduced the computation time to a couple of 
days. In addition, we added a module to estimate the orbital period of 
each particle before the final classification to ensure
that it is integrated for a fixed number of orbital periods rather than a 
fixed global integration time. The orbital periods vary significantly with
radius / binding energy for a typical galaxy: close to the centers we
find orbital periods less than 10~Myr and at 3~$R_\text{eff}$ they can be longer than 1~Gyr.

\begin{figure}
\centering
\includegraphics[width =
  0.23\textwidth]{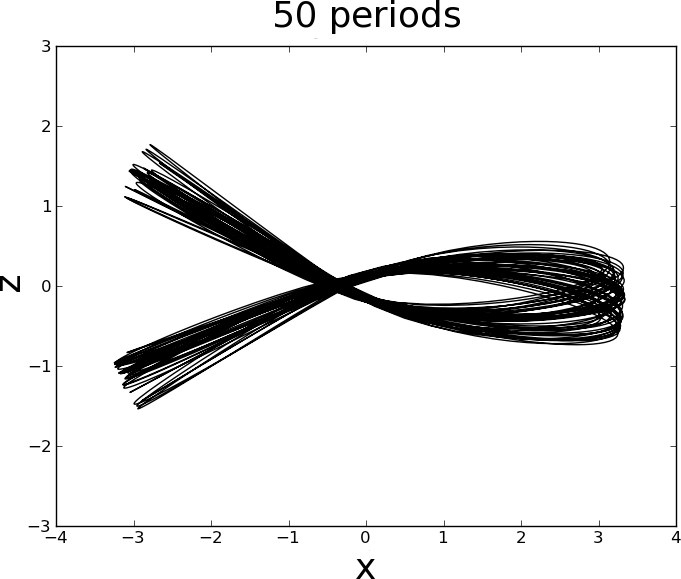}\ \includegraphics[width= 0.23\textwidth]{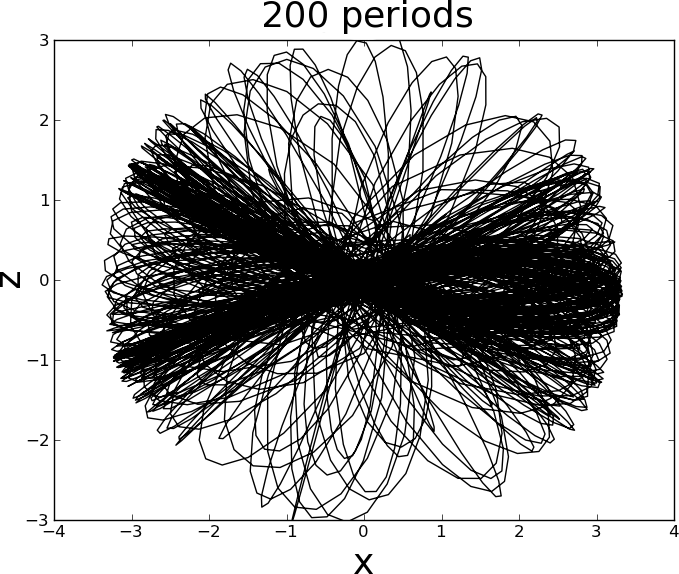} 
\caption{An example for a `sticky' orbit: A star particle of galaxy
  M0175 looks like a regular `fish orbit' (a 2:3-resonance, left
  panel) if integrated for 50 orbital periods ($\sim\!7.5$~Gyrs). After
  longer integration for 200 periods ($\sim\!30$~Gyrs) the orbit
  shows its irregular nature (right panel). For the classification in
  this paper all orbits are integrated for $\lesssim 50$ orbital
  periods.}  
\label{fig:sticky_orbit}
\end{figure}

The orbital period is usually defined as the inverse of the dominant
frequency along the major axis. To first order, the particle should
pass the $y$-$z$-plane twice for each period. To get an estimate for the
orbital period of each particle we first integrate (with low accuracy)
until the desired orbital periods sign changes (corresponding to
$\sim \! 40$ periods) have occurred. We then terminate the integration
and then re-integrate the orbit for the same time at high accuracy for the
orbit classification.

To find an optimal integration time, the trajectory has to be sufficiently 
long to detect peaks and their resonances in the Fourier spectra. On the 
other hand, too long integration time leads to inaccuracies and the 
sampling with the FFT can become to coarse to detect low amplitude 
frequencies. A complication are so-called `sticky orbits', irregular orbits 
that come near to a resonantly trapped region in phase space, stick to
it for some time and drift away slowly. These orbits look and behave 
very similarly to the resonant orbits nearby. Their actual irregular nature 
is not revealed until they have drifted away from the resonance (see 
Fig.~\ref{fig:sticky_orbit}), which in many cases requires integration for 
much more than the Hubble time. An appropriate integration time 
ensures that these orbits are classified as regular orbits. The 
classification is most robust if we integrate for about $40-50$ periods 
(which is shorter than the 100 periods as suggested by CA98 for a 
more detailed analysis). 

\subsection{Stability of the orbit classification}
\label{orbit_stability}

In this section we present test of how well the potential is sampled 
and how the sampling affects the orbit classification. For this we use the 
bootstrapping technique (see \citealp{1994ApJ...427..165H, 
2003ApJ...597..893N}). From the original ensemble of particles, 
we randomly draw the same number of particles with replacement. This 
is done separately for each particles species (gas, stars, and dark 
matter) to maintain the mass of each component.

\begin{figure}
\centering
\includegraphics[width =
  0.475\textwidth]{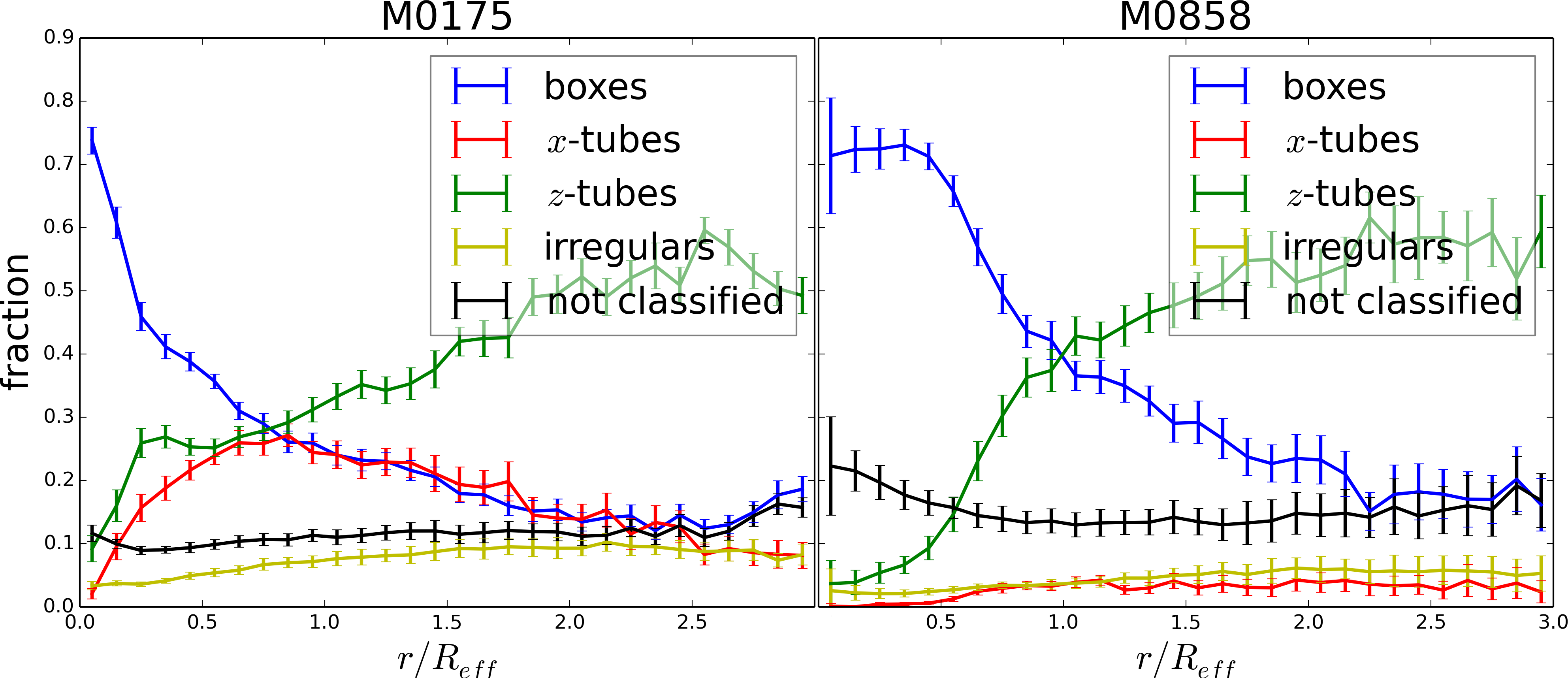} 
\caption{
  Orbit fractions as a function of radius for a bootstrapping ensemble of 
  realizations of two representative galaxies, M0175 and M0858. Solid 
  lines are the mean of 50 bootstrapped realizations the errorbars 
  indicate the standard deviations. The uncertainties due to the coarse 
  sampling of phase space are small.}
\label{fig:stability_bootstrap}
\end{figure}

We analyse 50 different realizations for two representative galaxies 
without figure rotation (M0175 and M0858). In 
Fig.~\ref{fig:stability_bootstrap} we show the mean orbit fractions and 
the variations (standard derivations) indicated by the error bars. The 
orbit classification is very robust and hardly affected by the sampling of 
phase space. The variations are typically small for the stellar 
component and can increase slightly at large radii ($r > 2 R_\text{eff}$; 
see \citealp{2014MNRAS.438.2701W} for the effect of time averaging 
the potential to measurements at large radii). Most importantly, none of 
the conclusion in the paper are affected by sampling issues.
The small variations in orbit fractions are caused by individual particles 
that change orbit families asymmetrically (see also 
Sec.~\ref{orbit_class}) for different realizations of the potential. Typically 
almost 90\% of the particles in the main orbit families do not change 
classification.

\begin{figure}
\centering
\includegraphics[width =
  0.475\textwidth]{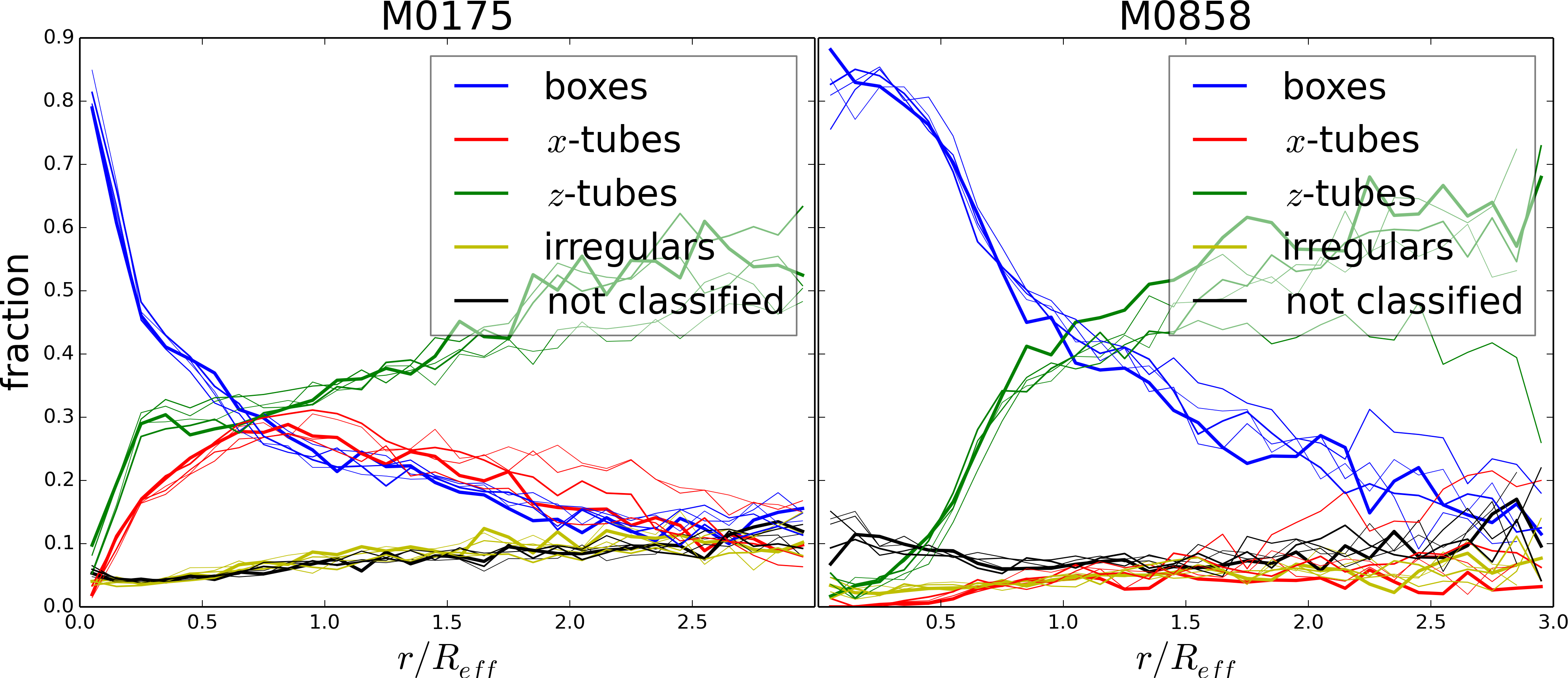} 
\caption{
  The time evolution of the orbit fractions as a function of radius for the 
  last five snapshots ($\sim \! 0.5$~Gyrs; the first snapshot is indicated 
  by a thick line) for the same galaxies as in 
  Fig.~\ref{fig:stability_bootstrap}. The fractions are stable with 
  moderate fluctuations at large radii ($r > 2 R_\text{eff}$).} 
\label{fig:stability_time}
\end{figure}

In Fig.~\ref{fig:stability_time} we investigate the evolution of orbit 
fractions in time for the last five snapshots covering about 500~Myrs. 
Again for both galaxies the orbit fractions are stable for the main part of 
the galaxies. At larger radii ($r > 2 R_\text{eff}$) the fluctuations with time 
become larger due to moving substructures (e.g. M0858).

In summery, the orbit fractions presented in section~\ref{results} are 
reliable and stable for a few dynamical times. Therefore none of the 
conclusions in the paper are affected by sampling issues and time 
evolution of the modelled galaxies.

\subsection{Figure rotation}

Since we have fixed the potential using a time-independent SCF potential, the classification procedure does not account for the possibility of a rotating figure.
Some of our galaxies, however, show clear signs of figure rotation.
We find that the systems that rotate are rather prolate or have bar-like structures that rotate.

The rotation of the figure is also imprinted in the motion of the stars.
Using the classification scheme without caution would lead to an incorrect orbit classification.
Hence, we excluded galaxies that show signs of figure rotation from our orbit analysis.
These are M0380, M0549, M0763, M0858, M1192 and M1306.

A direct integration of rotating figure into the CA98 classification scheme 
would be very demanding or even impossible, because a) the 
measurement of figure rotation is quite error-prone, b) the amount of 
figure rotation can change a lot over time due to external torques and c) 
figure rotation can be differential. The latter point means that the 
principal axis (of the reduced inertia tensor) change with radius and 
time. This not only complicates the methods of classification but a static 
potential is actually a premise of orbit theory. Hence, we restrict 
ourselves to galaxies with slowly varying/rotating potentials. In order to 
check the robustness of the classification in slowly rotating potentials, 
we added small amounts of rigid rotation to the SCF potentials and 
found that the orbit composition within a few effective radii does not 
change much for moderate pattern speeds. Therefore we trust the 
classification for all but the excluded galaxies.

\subsection{Line-of-sight kinematics}
\label{LOS_kin}

To connect the orbital structure of the galaxies to their
line-of-sight kinematics we present two-dimensional maps of the LOS
velocity, and dispersion as well as $h_3$ (see below). The maps are
constructed in a similar way to observations and all details can be
found in \citet{Naab_etal_2013}. Here we just give a short summary.
In a first step the galaxies are oriented along their principal axes
and projected along the intermediate axis, resulting in an edge-on
view. We split each stellar particle into 60 pseudo-particles with a
Gaussian distribution of standard deviation $0.3$~kpc. The
pseudo-particles are gathered into Voronoi bins such that each final 
bin has a signal-to-noise ratio of about 40 \citep{Cappellari_Copin_2003}.

For each bin the line-of-sight velocity profile $P(v)$ is fitted
\citep{2006A&A...445..513V} with Gauss-Hermite functions following
\cite{Gerhard_1993} and \cite{van_der_Marel_Franx_1993}, using terms
up to the fourth order:   
\begin{align}
P(v) = \frac{\gamma}{\sqrt{2 \pi \sigma^2}} \, \mathrm e^{-\frac{w^2}{2}}
  \, \Big( 1 + h_3 H_3(w) + h_4 H_4(w) \Big), \label{eq:gauss_hermite}
\end{align}
where
\begin{align}
w \equiv \frac{v - V_0}{\sigma}
\end{align}
is the normalized deviation of the velocity from the mean velocity $V_0$,
and $H_3$ and $H_4$ are Gauss-Hermite polynomials. $\gamma$ is a
normalization parameter and also one of the fitting parameters
$(\gamma, V_0, \sigma, h_3, h_4)$. The amplitude $h_3$ of the third
order term is connected to the skewness and $h_4$ is a measure of the
kurtosis of the velocity profile. Skewness and kurtosis, however, are
more sensitive to statistical deviations at the far ends of the wings
\citep{van_der_Marel_Franx_1993}. 

We also calculate the $\lambda_R$-parameter as introduced by
\cite{Emsellem_etal_2007, 2011MNRAS.414..888E}, a
luminosity weighted measure of rotation: 
\begin{align}
\lambda_R = \frac{\sum_{i=1}^{N} F_i \, r_i |V_{0,i}|}{
  \sum_{i=1}^{N} F_i \, r_i \sqrt{V_{0,i}^2 + \sigma_i^2}}, \label{eq:lambda_R}
\end{align}
where $F_i$ is the flux (which here is the mass of the stars, for a 
constant light-to-mass ratio) in bin $i$ of the projected galaxy, $r_i$ 
is its projected radius, $V_{0,i}$ is the line-of-sight velocity and 
$\sigma_i$ is the line-of-sight velocity dispersion of bin $i$.

\section{Results} 
\label{results}

\begin{table*}
\centering
\begin{tabular}{c|ccccc|ccc|c}
\multirow{2}{*}{ID} & $M_*$ & $R_\text{eff}$ & \multirow{2}{*}{triax. $T$} & \multirow{2}{*}{$\lambda_R$} & {\it in situ} & box & $z$-tube & $x$-tube & Eff. prograde \\
  & [$10^{10}\,M_\Sun$] & [kpc] &   &   & frac. & frac. & frac. & frac. & $z$-tube frac. \\
\hline
M0040 & 49.98 & 12.92 & 0.69 & 0.13 & 0.11 & 0.44 & 0.24 & 0.12 & 0.04 \\
M0053 & 69.45 & 13.03 & 0.51 & 0.093 & 0.19 & 0.29 & 0.40 & 0.17 & 0.02 \\
M0069 & 49.40 &  8.84 & 0.68 & 0.15 & 0.15 & 0.43 & 0.22 & 0.19 & 0.09 \\
M0089 & 52.34 & 10.43 & 0.45 & 0.074 & 0.084 & 0.44 & 0.27 & 0.11 & 0.03 \\
M0094 & 47.90 &  7.53 & 0.54 & 0.098 & 0.16 & 0.38 & 0.39 & 0.13 & 0.12 \\
M0125 & 43.35 &  9.07 & 0.85 & 0.078 & 0.12 & 0.26 & 0.17 & 0.48 & 0.02 \\
M0162 & 36.44 &  9.78 & 0.80 & 0.074 & 0.081 & 0.52 & 0.19 & 0.13 & 0.06 \\
M0163 & 35.20 & 10.39 & 0.63 & 0.31 & 0.098 & 0.49 & 0.25 & 0.10 & 0.12 \\
M0175 & 36.79 &  7.37 & 0.58 & 0.058 & 0.14 & 0.34 & 0.35 & 0.19 & 0.01 \\
M0190 & 31.48 &  6.98 & 0.77 & 0.083 & 0.093 & 0.51 & 0.18 & 0.17 & 0.01 \\
M0204 & 26.85 &  6.49 & 0.37 & 0.10 & 0.12 & 0.31 & 0.45 & 0.12 & 0.06 \\
M0209 & 19.96 &  3.81 & 0.41 & 0.14 & 0.17 & 0.46 & 0.36 & 0.06 & 0.09 \\
M0215 & 27.64 &  5.04 & 0.49 & 0.14 & 0.16 & 0.33 & 0.32 & 0.25 & 0.10 \\
M0224 & 24.84 &  5.95 & 0.26 & 0.16 & 0.14 & 0.24 & 0.54 & 0.10 & 0.11 \\
M0227 & 30.90 &  7.91 & 0.61 & 0.24 & 0.10 & 0.35 & 0.38 & 0.16 & 0.18 \\
M0259 & 19.83 &  4.53 & 0.06 & 0.40 & 0.15 & 0.28 & 0.58 & 0.02 & 0.30 \\
M0290 & 22.03 &  3.57 & 0.10 & 0.48 & 0.19 & 0.18 & 0.73 & 0.03 & 0.47 \\
M0300 & 18.65 &  4.58 & 0.37 & 0.19 & 0.12 & 0.42 & 0.40 & 0.05 & 0.17 \\
M0305 & 25.76 &  8.92 & 0.46 & 0.12 & 0.11 & 0.50 & 0.21 & 0.11 & 0.01 \\
M0329 & 21.33 &  4.32 & 0.49 & 0.071 & 0.16 & 0.44 & 0.29 & 0.16 & 0.03 \\
M0380 & 17.08 &  4.04 & 0.57 & 0.46 & 0.17 & {\color{gray} 0.44} & {\color{gray} 0.34} & {\color{gray} 0.10} & {\color{gray} 0.15} \\
M0408 & 17.71 &  3.60 & 0.07 & 0.37 & 0.20 & 0.23 & 0.67 & 0.02 & 0.34 \\
M0443 & 23.08 &  2.74 & 0.44 & 0.088 & 0.24 & 0.37 & 0.42 & 0.14 & 0.11 \\
M0501 & 16.31 &  4.29 & 0.55 & 0.075 & 0.16 & 0.47 & 0.23 & 0.15 & 0.05 \\
M0549 & 11.64 &  4.66 & 0.33 & 0.46 & 0.17 & {\color{gray} 0.44} & {\color{gray} 0.31} & {\color{gray} 0.09} & {\color{gray} 0.16} \\
M0616 & 13.04 &  4.07 & 0.72 & 0.077 & 0.16 & 0.47 & 0.19 & 0.21 & 0.02 \\
M0664 & 10.39 &  2.91 & 0.64 & 0.14 & 0.17 & 0.33 & 0.34 & 0.23 & 0.00 \\
M0721 & 13.35 &  2.32 & 0.14 & 0.40 & 0.38 & 0.29 & 0.60 & 0.02 & 0.24 \\
M0763 & 13.68 &  4.28 & 0.30 & 0.32 & 0.099 & {\color{gray} 0.39} & {\color{gray} 0.43} & {\color{gray} 0.06} & {\color{gray} 0.25} \\
M0858 & 14.26 &  2.69 & 0.31 & 0.52 & 0.32 & {\color{gray} 0.50} & {\color{gray} 0.35} & {\color{gray} 0.03} & {\color{gray} 0.18} \\
M0908 & 13.43 &  2.83 & 0.06 & 0.44 & 0.38 & 0.20 & 0.69 & 0.01 & 0.38 \\
M0948 &  9.23 &  4.75 & 0.69 & 0.088 & 0.12 & 0.31 & 0.20 & 0.35 & 0.03 \\
M0959 &  8.41 &  2.94 & 0.88 & 0.091 & 0.18 & 0.31 & 0.13 & 0.46 & 0.03 \\
M0977 &  6.32 &  3.26 & 0.47 & 0.35 & 0.37 & 0.47 & 0.26 & 0.11 & 0.16 \\
M1017 &  8.87 &  2.27 & 0.61 & 0.084 & 0.29 & 0.46 & 0.33 & 0.12 & 0.04 \\
M1061 &  7.20 &  3.13 & 0.64 & 0.066 & 0.21 & 0.54 & 0.16 & 0.17 & 0.02 \\
M1071 & 10.82 &  2.37 & 0.50 & 0.14 & 0.21 & 0.32 & 0.40 & 0.20 & 0.12 \\
M1091 & 10.46 &  2.00 & 0.54 & 0.040 & 0.25 & 0.57 & 0.25 & 0.09 & 0.02 \\
M1167 & 10.24 &  2.31 & 0.54 & 0.062 & 0.26 & 0.55 & 0.25 & 0.11 & 0.07 \\
M1192 &  6.05 &  2.64 & 0.69 & 0.50 & 0.20 & {\color{gray} 0.54} & {\color{gray} 0.18} & {\color{gray} 0.16} & {\color{gray} 0.11} \\
M1196 & 10.74 &  3.04 & 0.11 & 0.45 & 0.33 & 0.27 & 0.61 & 0.01 & 0.41 \\
M1306 &  9.04 &  1.84 & 0.13 & 0.59 & 0.28 & {\color{gray} 0.39} & {\color{gray} 0.50} & {\color{gray} 0.03} & {\color{gray} 0.31} \\
\end{tabular}
\caption{Properties of simulated galaxies: $M_*$ is the stellar mass
  of the galaxy, i.e.\ the stellar mass within $10\% \, R_{200}$,
  $R_\text{eff}$ is the half mass radius of the galaxy's stars
  (i.e.\ the effective radius assuming a constant mass-to-light ratio
  and ignoring gas), the triaxiality $T$ is measured at
  $R_\text{eff}$, $\lambda_R$ is the rotational parameter as defined
  in formula~\ref{eq:lambda_R}, the {\it in situ} fraction is the fraction
  of the stars that were formed {\it in situ} since $z=2$ and the orbit
  class fractions were determined with the modified CA98 classifier
  within $3 R_\text{eff}$. The effective prograde $z$-tube fraction is
  defined as the difference between the prograde and the retrograde
  $z$-tube fraction. Uncertain orbit fractions for galaxies with significant 
  figure rotation are given in grey.}
\label{tab:gal_props}
\end{table*}

\subsection{Orbit population and triaxiality}
\label{orbit_pop}

\begin{figure*}
\centering
\includegraphics[width = 0.325\textwidth]{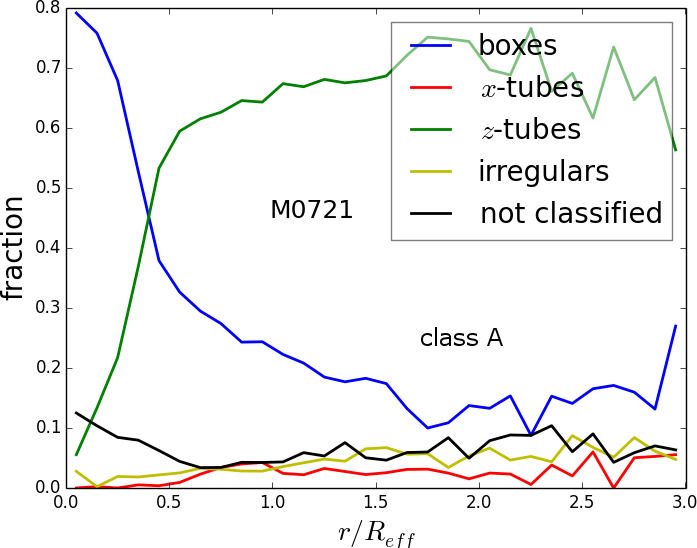}
\includegraphics[width = 0.325\textwidth]{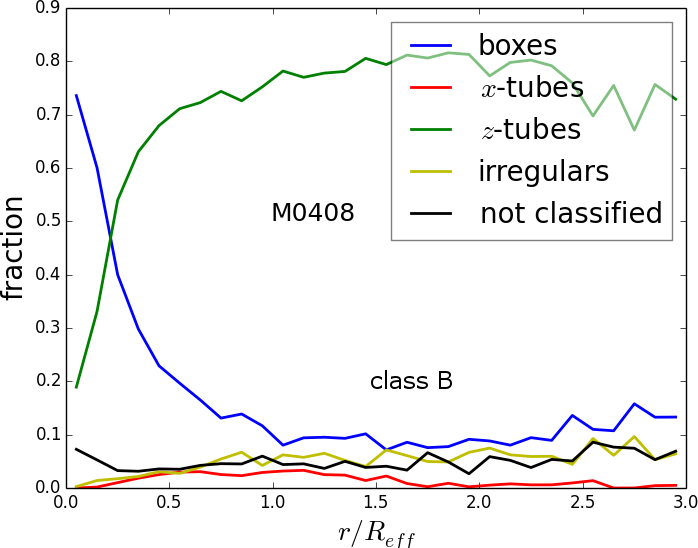}
\includegraphics[width = 0.325\textwidth]{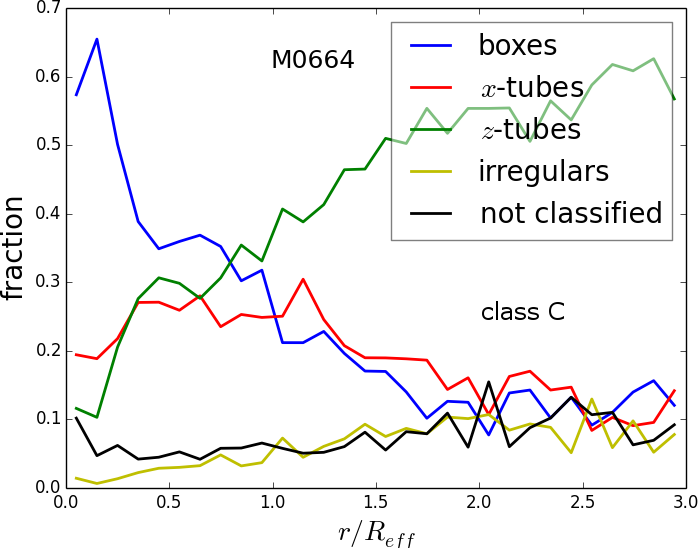}
\vspace{0.05cm}

\includegraphics[width = 0.325\textwidth]{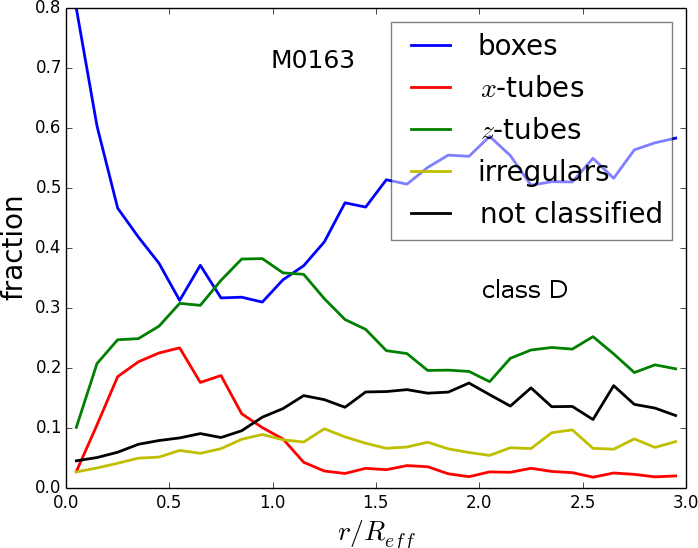}
\includegraphics[width = 0.325\textwidth]{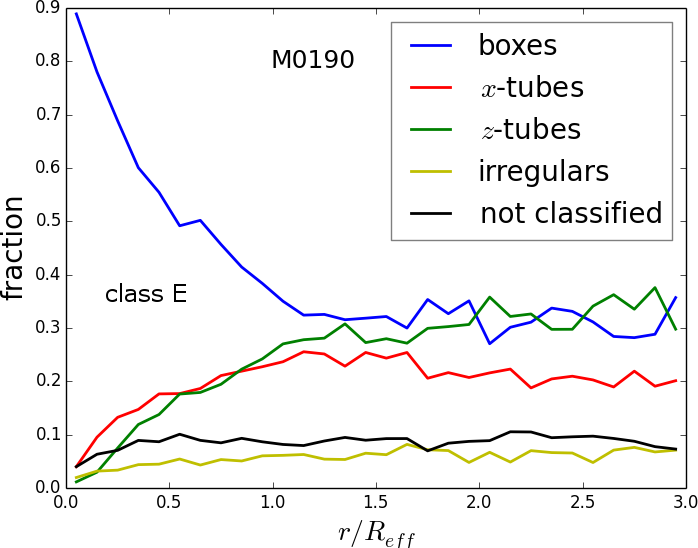}
\includegraphics[width = 0.325\textwidth]{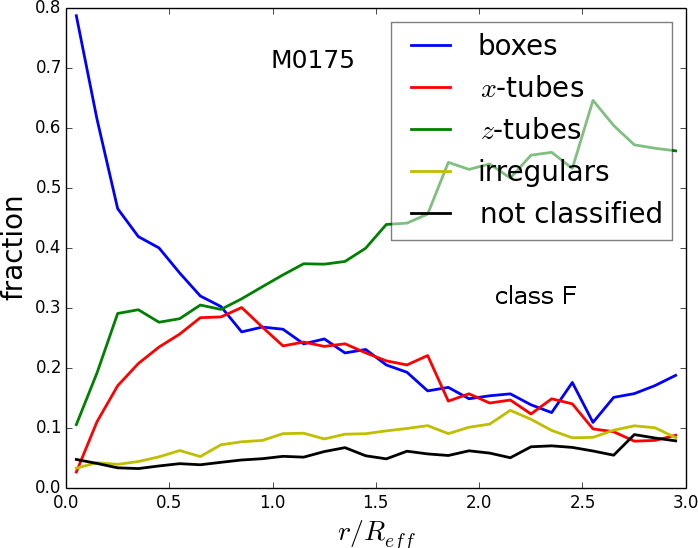}
\caption{Fraction of box (blue), $x$-tube (red), $z$-tube (green),
  irregular (light green) and not classified (black) orbits as a
  function of radius for six galaxies with different formation
  histories \citep{Naab_etal_2013}. All simulated galaxies are
  dominated by box orbits at the center (which probably is an 
  artifact of the simulation model we use). Galaxies of classes A and B
  besom $z$-tube dominated at $\sim \! 0.5 R_{\mathrm{eff}}$, $x$-tubes can
  be found for classes C, D, E, and F. Typically less than 10 per cent
  of all orbits are irregular or not classified.}
\label{fig:orbitstructs_repr}
\end{figure*}

\begin{figure*}
\centering
\includegraphics[width = 0.31\textwidth]{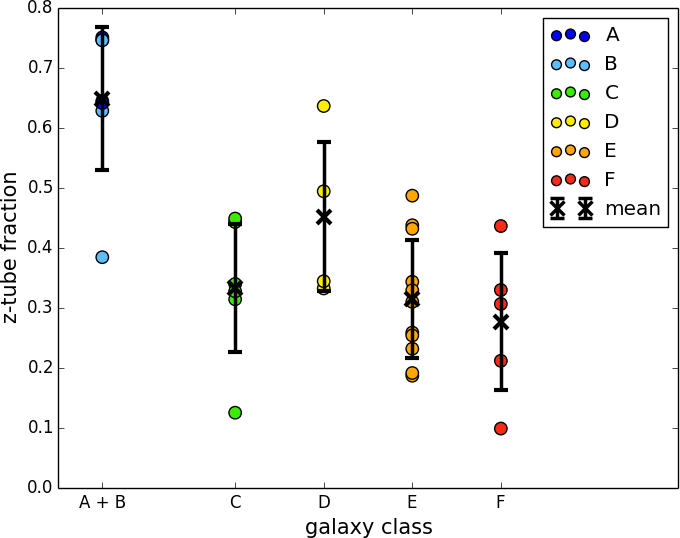} \quad
\includegraphics[width = 0.31\textwidth]{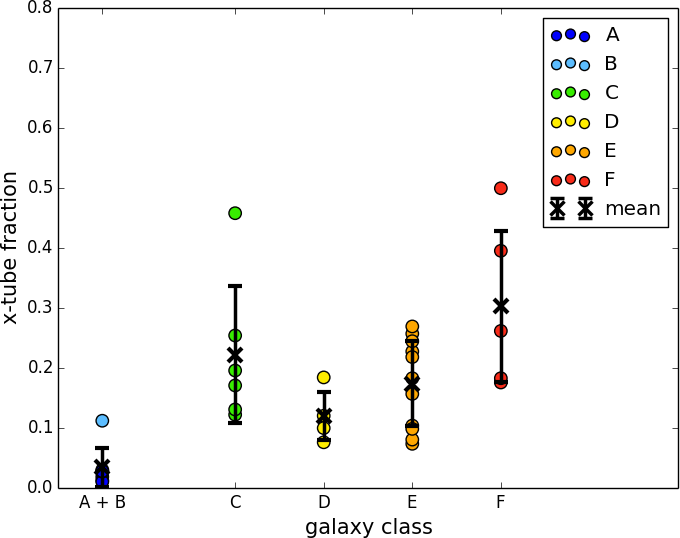} \quad
\includegraphics[width = 0.31\textwidth]{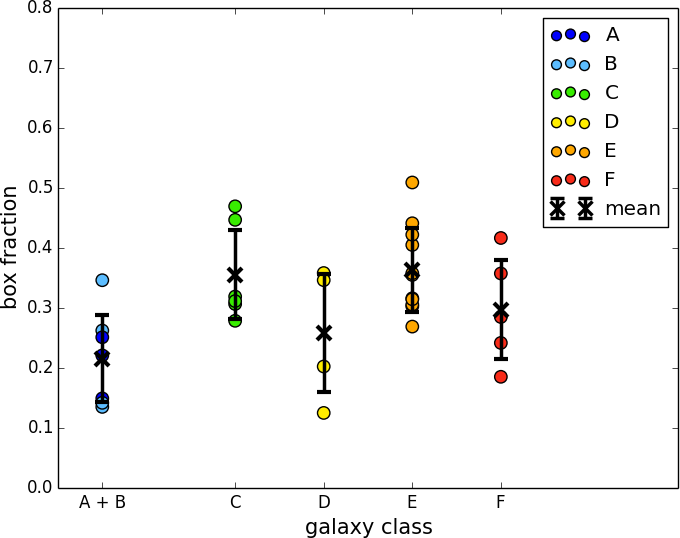} \quad
\caption{Fraction of $z$-tubes (left), $x$-tubes (middle), and box orbits 
   (right) in the radial range of $0.5 \, R_\text{eff}$ to $1.5 \, R_\text{eff}$ 
   (smaller radii are dominated by box orbits---which probably is an 
   artefact of the simulation model we use) for different  assembly 
   classes as defined in \citet{Naab_etal_2013}. For the individual 
   classes we also show the mean fractions with their standard 
   deviation as error bars. Fast rotators (classes A,B, and D) have 
   high $z$-tube fractions, slow rotators (C, E, and F) have significant 
   $x$-tube and box orbit contributions.}   
\label{fig:galclass_trend}
\end{figure*}

\begin{figure*}
\centering
\includegraphics[width = 0.32\textwidth]{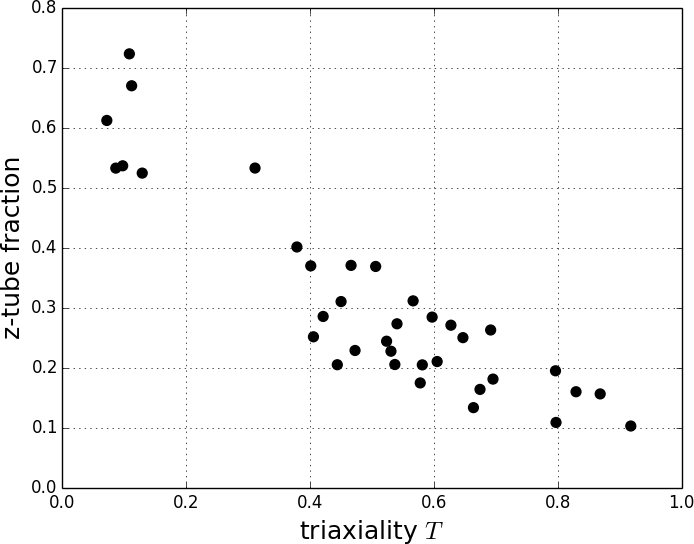}
\includegraphics[width = 0.32\textwidth]{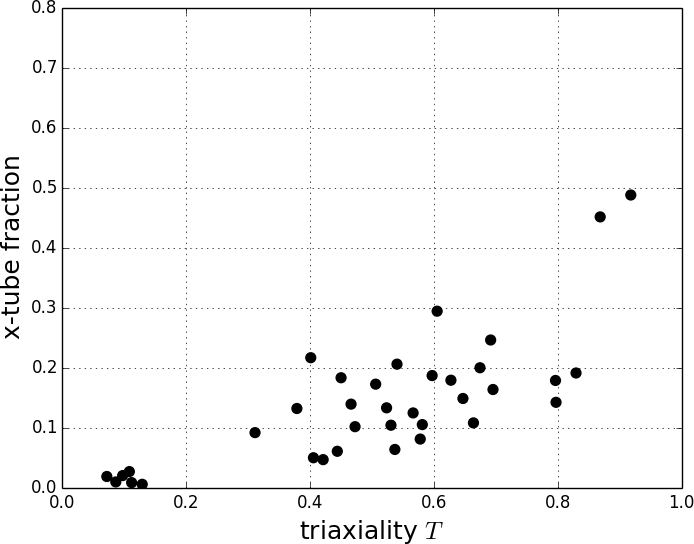}
\includegraphics[width = 0.32\textwidth]{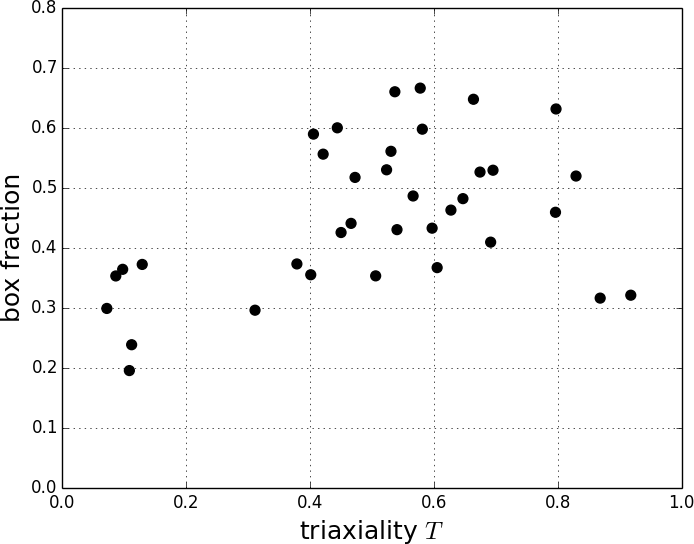}
\caption{The total fraction of stars on $z$-tubes (left panel), $x$-tubes 
  (middle panel), and box orbits (right panel) as a function of galaxy 
  stellar triaxiality $T$ (both measured within $R_{\mathrm{eff}}$). 
  Oblate systems ($T=0$) are $z$-tube dominated, prolate systems 
  ($T=1$) can support $x$-tubes, and triaxial systems ($T=0.5$) have the 
  highest box orbit fraction as predicted by theory \citep{Statler_1987} 
  found for binary merger remnants \citep{2005MNRAS.360.1185J}.}
\label{fig:orbfrac_T}
\end{figure*}

\begin{figure*}
\centering
\includegraphics[width = 0.26\textwidth]{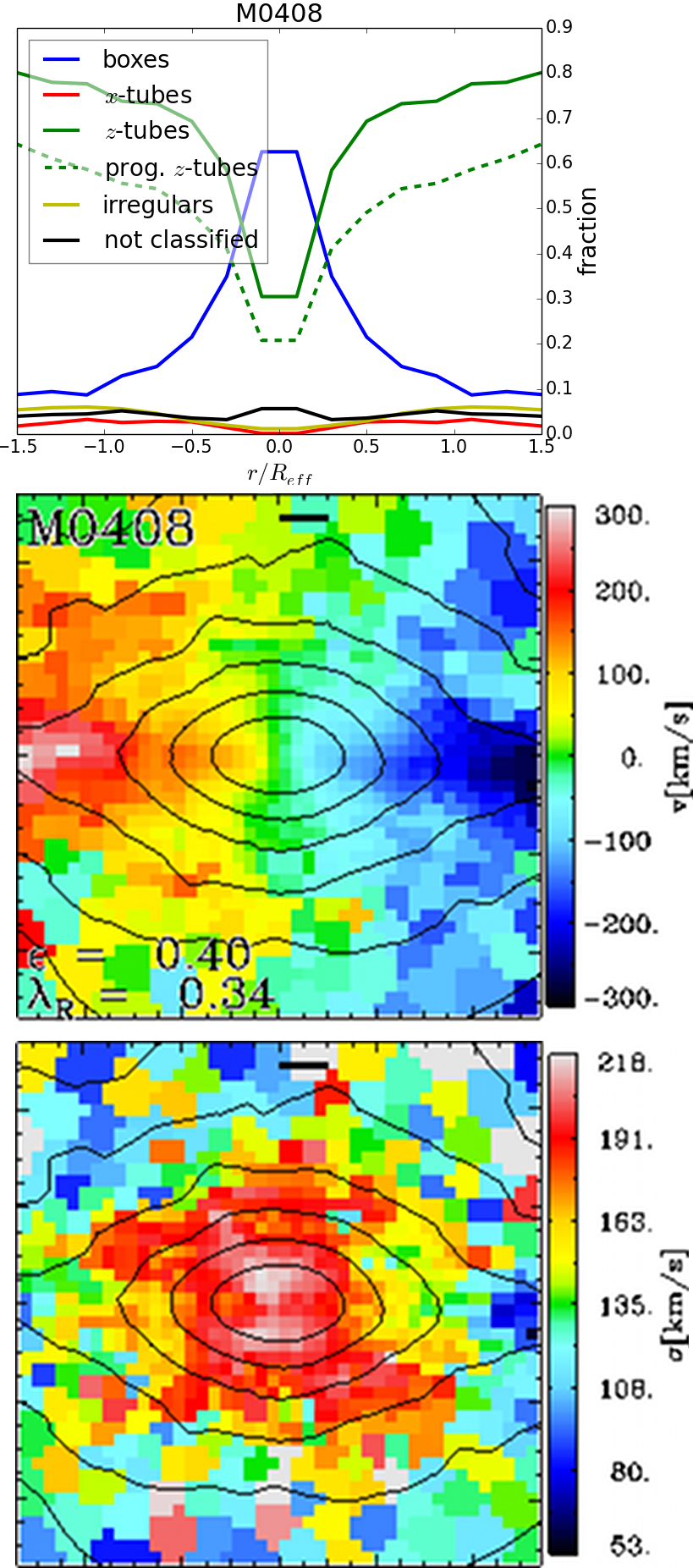}
\includegraphics[width = 0.26\textwidth]{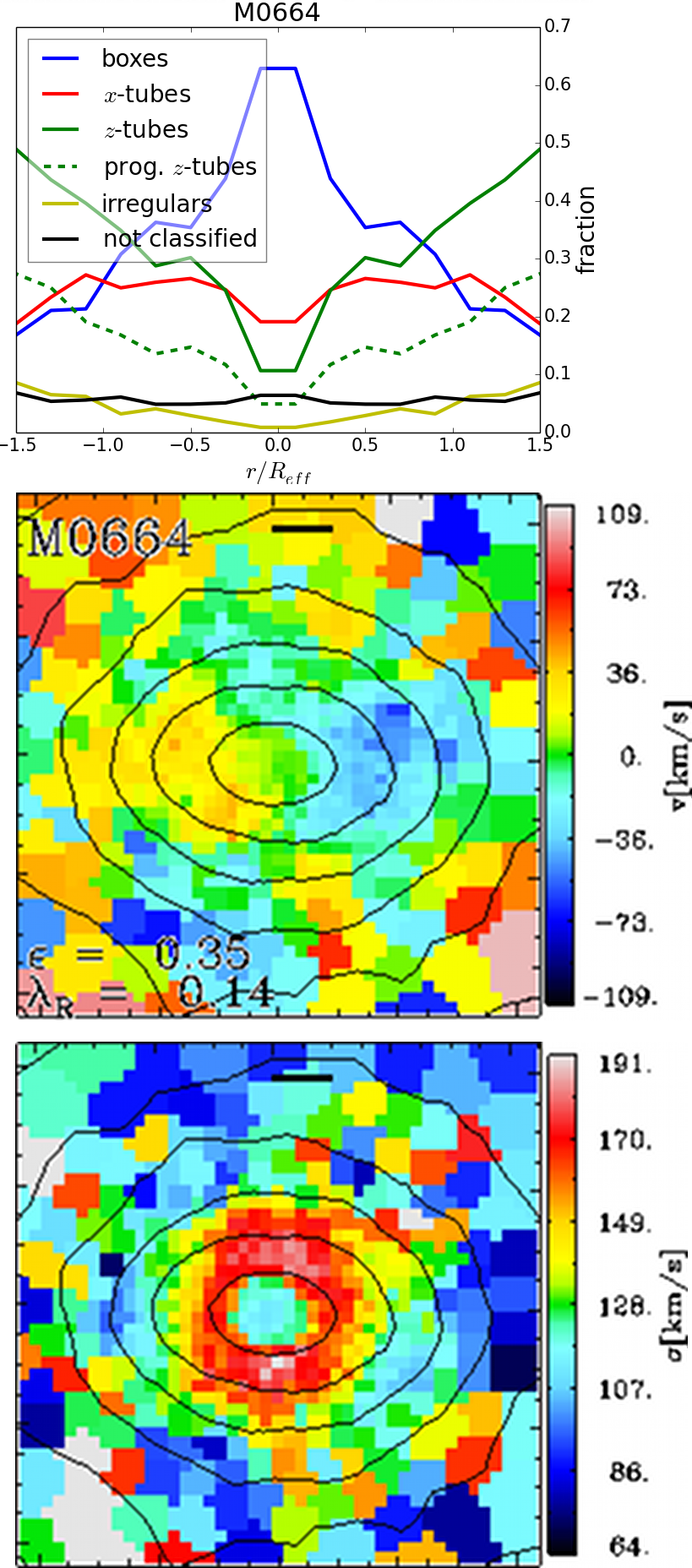}
\includegraphics[width = 0.26\textwidth]{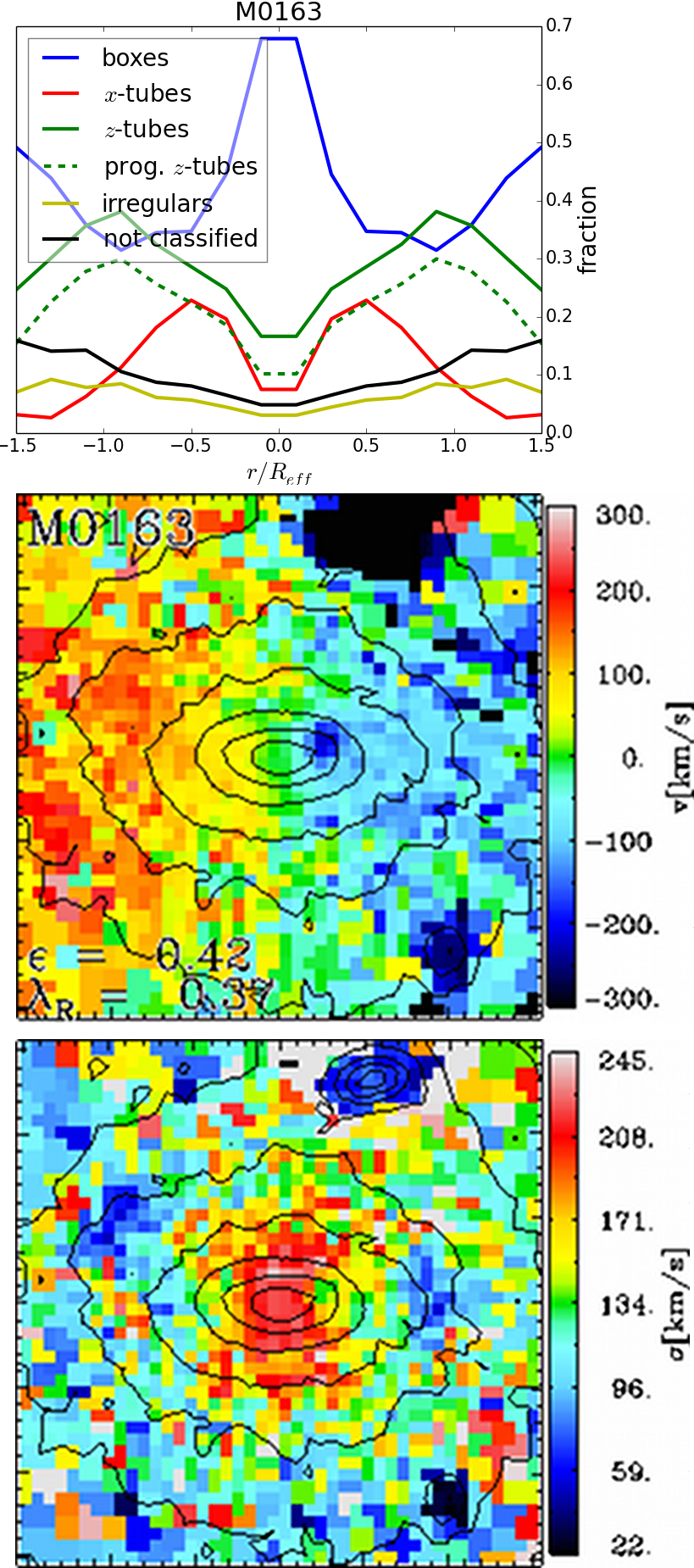}
\vspace{0.5cm}

\includegraphics[width = 0.26\textwidth]{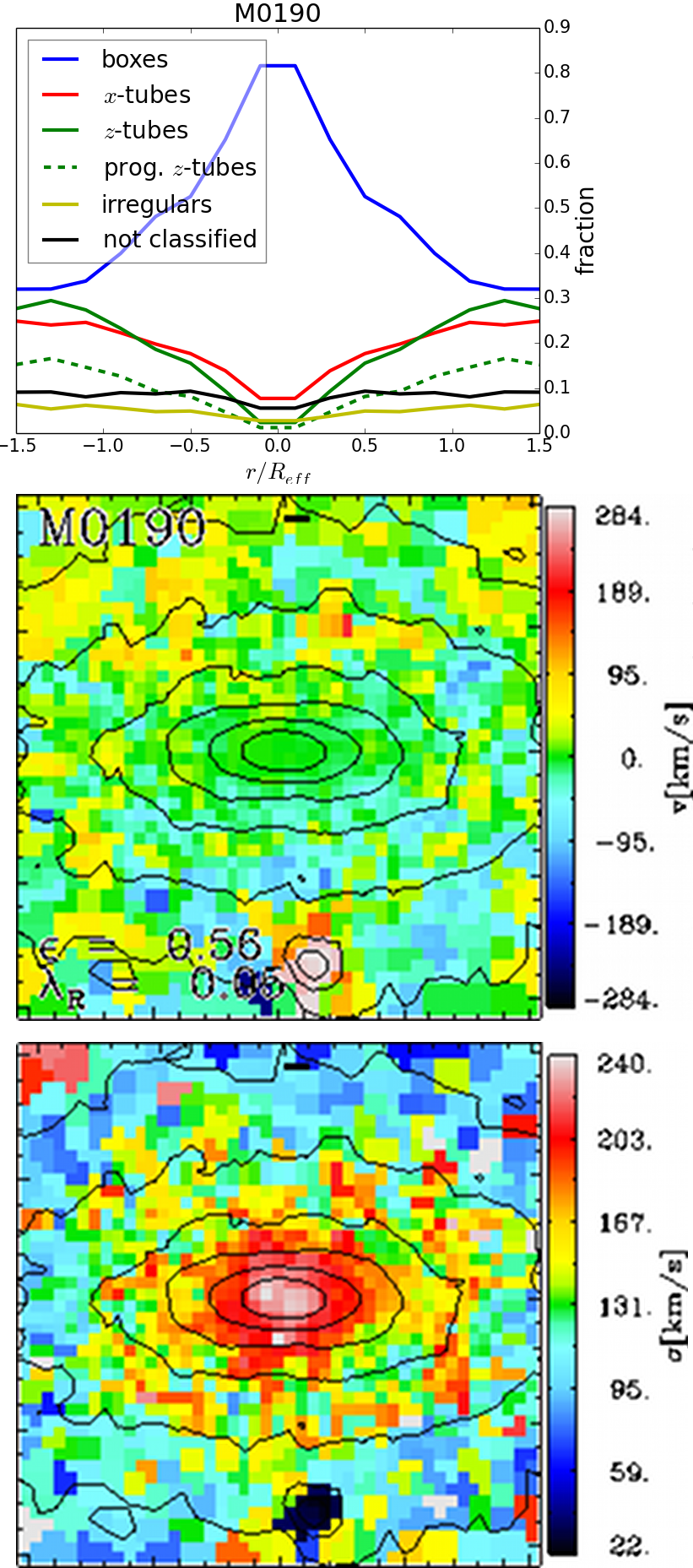}
\includegraphics[width = 0.26\textwidth]{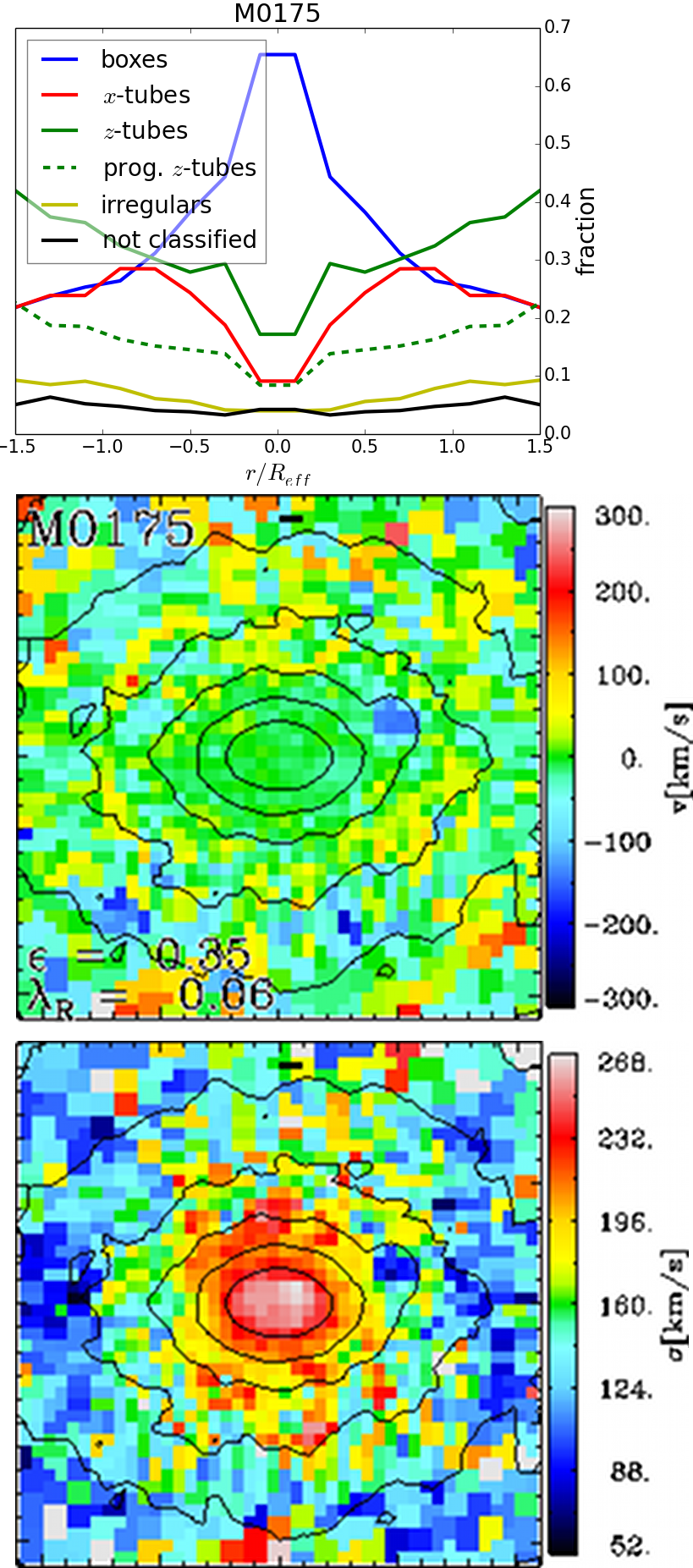}
\includegraphics[width = 0.26\textwidth]{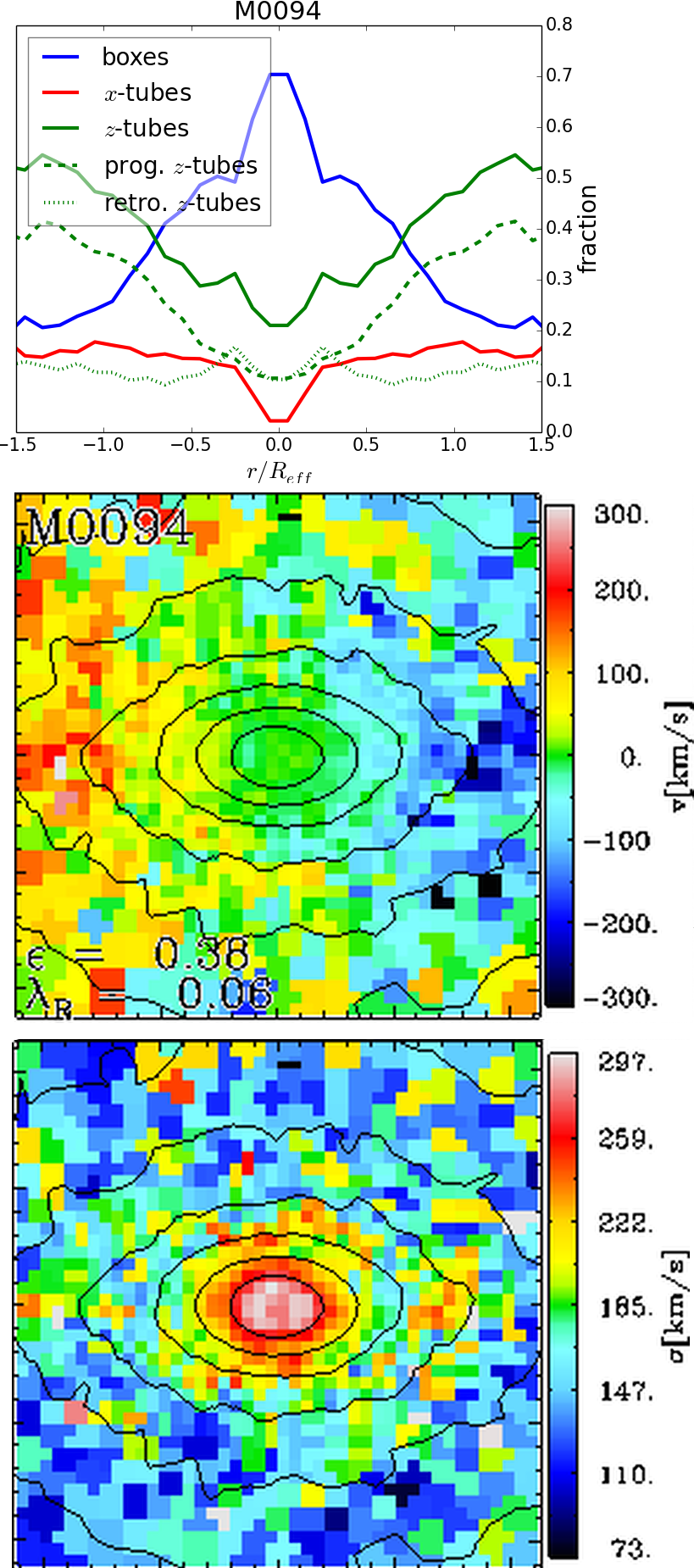}
\caption{Two-dimensional line-of-sight velocity and velocity
  dispersion maps with isodensity contours for the six prototypical
  galaxies within $1.5\,R_\mathrm{eff}$. The $\lambda_R$ parameter and
  the isophotal ellipticity at $R_{\mathrm{eff}}$ is given in the
  line-of-sight velocity panels. The black bar indicates 1~kpc. For
  comparison we show the radial orbit fractions
  (Fig.~\ref{fig:orbitstructs_repr}) mirrored at $r = 0$. Galaxies
  with fastest rotation (M0408 and M0163), have the 
  highest prograde $z$-tube factions with relatively few canceling 
  retrograde $z$-tubes. If the latter is not the case, high $z$-tube 
  fractions do not yield strong rotation (M0664 and M0175). M0094
  has a counter-rotating core, where we see both: at the core
  the retrograde $z$-tubes (slightly) dominate and at larger radii the
  prograde $z$-tubes (clearly) dominate.}
\label{fig:orbits_LOSVD}
\end{figure*}

We classify the orbits for stellar and dark matter particles up to
3 effective radii for all 42 galaxies {(a summary is given in Tab.~\ref{tab:gal_props}). The particles populate all major orbit classes:
box orbits and boxlets (boxes), $z$-tubes (minor-axis loop orbits), inner and
outer $x$-tubes (major-axis loop orbits), and irregular orbits (see Fig.~\ref{fig:classes_traj}). In general, the dominant
orbit families  are boxes and $z$-tubes, sometimes $x$-tubes, with 
significant radial variation. A small fraction of the
orbits, typically less than $\sim 10\%$, are irregular or cannot be
classified. 

In Fig.~\ref{fig:orbitstructs_repr}  we show the radial orbit 
distribution for six galaxies that belong to distinct formation
classes as discussed in \citet{Naab_etal_2013}.  M0721 (upper left
panel in Fig.~\ref{fig:orbitstructs_repr}) is a fast rotating galaxy 
with gas-rich minor mergers only (class A) and M0408 (upper middle 
panel in Fig.~\ref{fig:orbitstructs_repr}) is a fast rotating remnant 
of a major merger with dissipation (class B).
Both are dominated by $z$-tubes, apart from the very center 
($r \lesssim 0.25 R_{\mathrm{eff}}$). Such an orbit structure is very 
generic for galaxies of class A and B and it is very similar to 
gas-rich major mergers (see e.g. 
\citealp{2006MNRAS.372..839N, 2010ApJ...723..818H}).

A slowly rotating gas-rich merger remnant (M0664, class C) is shown in the upper right panel of Fig.~\ref{fig:orbitstructs_repr}.
This galaxy is triaxial,
almost prolate ($T \simeq 0.6-0.7$), with a significant
population of major axis tubes inside one effective radius. Again,
this galaxy is dominated by minor-axis  tubes at large radii. M0163 
is a fast rotating remnant of a gas-poor major merger (class D, bottom
left panel of Fig.~\ref{fig:orbitstructs_repr}) and is dominated by
box orbits, in particular at the center and at large radii (though the box 
orbits in the outer parts are mostly from a substructures there). 
An example for a slowly rotating remnant of a gas-poor major merger
(class E) is shown in the middle bottom panel of
Fig.~\ref{fig:orbitstructs_repr}. M0190 is again dominated by box  
orbits at the center, however, at large radii box orbits, $z$-tubes
and $x$-tubes contribute equally. M0175, a massive galaxy
assembled by minor mergers since $z \approx 2$ (class E), is dominated
by boxes at the center and minor-axis tubes in the outer parts with a
significant majors-axis tube contribution at all radii (bottom right
panel of Fig.~\ref{fig:orbitstructs_repr}).

For all fast rotating galaxies of classes A and B (like M0721 and 
M0408) the orbit distributions are generic. All other galaxy classes 
show significant variations in the orbit distributions. For example 
not all galaxies of class D have such high box fraction at large radii 
as M0163. Still, we do find trends, in particular around the effective 
radius, $0.5 \lesssim R_{\mathrm{eff}} \lesssim 1.5$ (see
Fig.~\ref{fig:galclass_trend}). Fast rotators (classes A, B, and D)
tend to have more $z$-tubes than the other classes, in particular  
classes A and B. In slow rotators (classes C, E, and F) box orbits are
more abundant. Also, $x$-tubes are rare in fast rotators. However, a high 
$z$-tube fraction is not the only factor determining the amount of rotation 
as we will show in section~\ref{orbits_LOSVD}. A common feature for 
all galaxies of the radial orbit distributions is a very box orbit dominated 
center (typically more than $60\%$ are boxes). This can probably be 
ascribed to the simulation model we use as we will discuss later.

It has been shown that the orbital composition (orbit fractions) correlate 
with the three-dimensional shape (triaxiality) of the system 
\citep{Statler_1987, 2005MNRAS.360.1185J}. In Fig.~\ref{fig:orbfrac_T} we 
show $z$-tube, $x$-tube, and box orbit fractions as a function of triaxiality 
of the galaxies. As expected, $z$-tubes preferentially live in oblate 
($T \approx 0$) systems, where they can rotate around a well-defined 
minor axis, which is poorly defined in a prolate ($T \approx 1$) system. 
In a galaxy with $T = 1$ exactly there are no $z$-tubes at all.
$x$-tubes are most abundant in prolate systems with a well-defined major 
axis ($x$-axis) and the abundance of box orbits peaks roughly at the most 
triaxial systems ($T \approx 0.5$, see right panel of 
Fig.~\ref{fig:orbfrac_T}). The fact that all systems have at least $20\%$ 
box orbits indicates that none of the systems is perfectly oblate.

\subsection{Orbits and line-of-sight kinematics}
\label{orbits_LOSVD}

\begin{figure*}
\centering
\textbf{M0408}

\begin{minipage}{0.47\textwidth}
\centering
all particles\vspace{0.1cm}

\includegraphics[width = 0.99\textwidth]{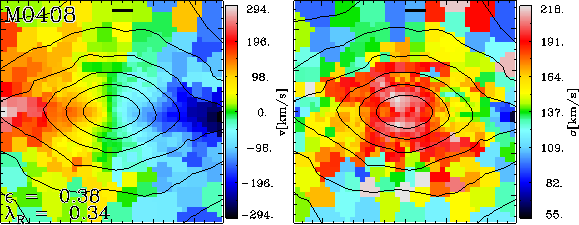}
\end{minipage} $\quad$
\begin{minipage}{0.47\textwidth}
\centering
boxes ($23.9\%$)\vspace{0.1cm}

\includegraphics[width = 0.99\textwidth]{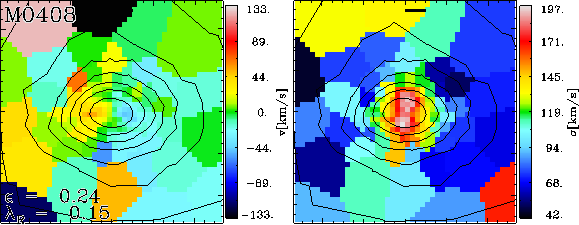}
\end{minipage}
\vspace{0.25cm}

\begin{minipage}{0.47\textwidth}
\centering
$z$-tubes ($66.4\%$)\vspace{0.1cm}

\includegraphics[width = 0.99\textwidth]{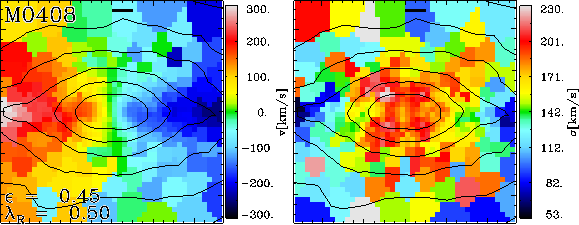}
\end{minipage} $\quad$
\begin{minipage}{0.47\textwidth}
\centering
$x$-tubes ($1.9\%$)\vspace{0.1cm}

\includegraphics[width = 0.99\textwidth]{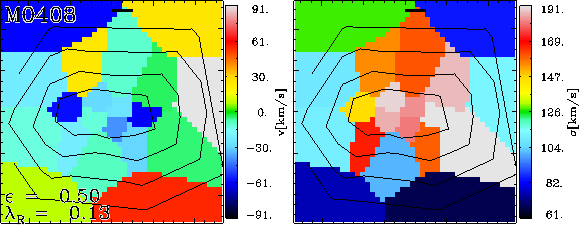}
\end{minipage}
\caption{The line-of-sight velocity and velocity dispersion maps with
  isodensity contours of all star particles of M0408 up to
  $1.5\,R_\text{eff}$ (top left),  for all stars on box orbits ($\sim
  24\%$, top right), stars on $z$-tubes ($\sim 66\%$, bottom left) and
  $x$-tubes ($\sim 2\%$, bottom right). The given orbit fractions are
  for stars in the maps and can differ from the values those given
  in Tab.~\ref{tab:gal_props}.  The rotation (see also the respective
  $\lambda_R$ values) and the flattening originates $z$-tubes. The high
  dispersion at the center is supported by box orbits.} 
\label{fig:LOSVD_decomp_M0408}
\end{figure*}

\begin{figure*}
\centering
\textbf{M0175}

\begin{minipage}{0.47\textwidth}
\centering
all particles\vspace{0.1cm}

\includegraphics[width = 0.99\textwidth]{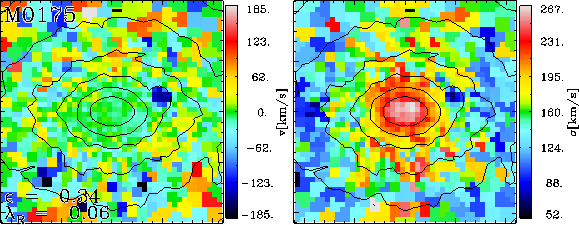}
\end{minipage} $\quad$
\begin{minipage}{0.47\textwidth}
\centering
boxes ($36.4\%$)\vspace{0.1cm}

\includegraphics[width = 0.99\textwidth]{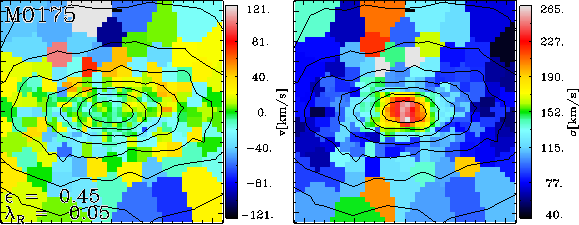}
\end{minipage}
\vspace{0.25cm}

\begin{minipage}{0.47\textwidth}
\centering
$z$-tubes ($32.9\%$)\vspace{0.1cm}

\includegraphics[width = 0.99\textwidth]{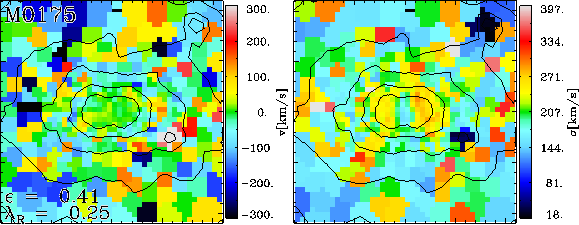}
\end{minipage} $\quad$
\begin{minipage}{0.47\textwidth}
\centering
$x$-tubes ($20.2\%$)\vspace{0.1cm}

\includegraphics[width = 0.99\textwidth]{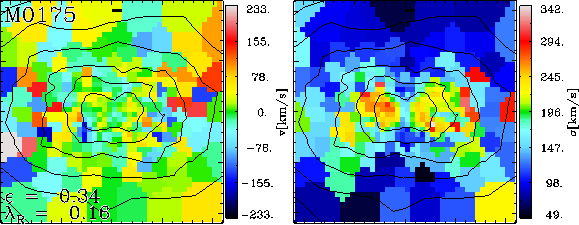}
\end{minipage}
\caption{The line-of-sight velocity and velocity dispersion maps of
  M0175 and its orbital components like for M0408 in
  Fig.~\ref{fig:LOSVD_decomp_M0408}. The particles on $z$-tubes (lower 
  left panel) are distributed equally on prograde and retrograde orbits
  resulting in low net rotation ($\lambda_R \sim 0.25$). The high
  dispersion inner part of the system is determined by the properties
  of box orbits (upper right panel). The system has a significant
  $x$-tube population contributing to the dispersion at $\sim 1
  R_{\mathrm{eff}}$.} 
\label{fig:LOSVD_decomp_M0175}
\end{figure*}

\begin{figure*}
\centering
\textbf{M0094}

\begin{minipage}{0.47\textwidth}
\centering
all particles\vspace{0.1cm}

\includegraphics[width = 0.99\textwidth]{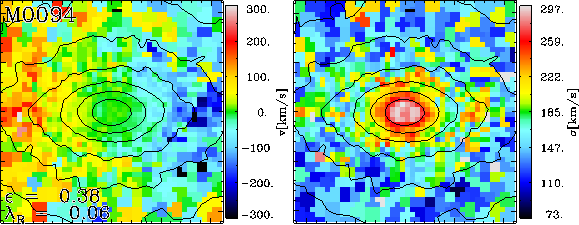}
\end{minipage} $\quad$
\begin{minipage}{0.47\textwidth}
\centering
boxes ($41.9\%$)\vspace{0.1cm}

\includegraphics[width = 0.99\textwidth]{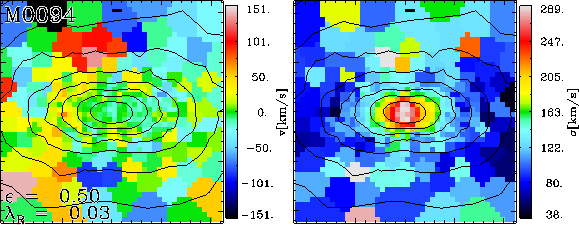}
\end{minipage}
\vspace{0.25cm}

\begin{minipage}{0.47\textwidth}
\centering
$z$-tubes ($35.6\%$)\vspace{0.1cm}

\includegraphics[width = 0.99\textwidth]{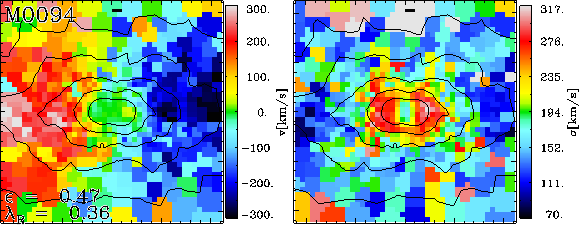}
\end{minipage} $\quad$
\begin{minipage}{0.47\textwidth}
\centering
$x$-tubes ($14.0\%$)\vspace{0.1cm}

\includegraphics[width = 0.99\textwidth]{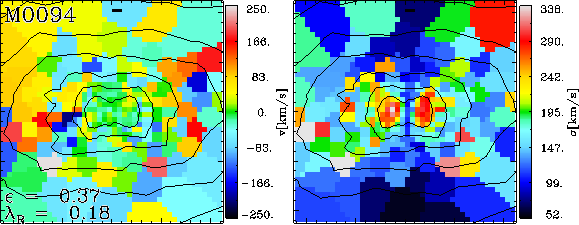}
\end{minipage}
\vspace{0.25cm}

\begin{minipage}{0.47\textwidth}
\centering
prograde $z$-tubes ($22.7\%$)\vspace{0.1cm}

\includegraphics[width = 0.99\textwidth]{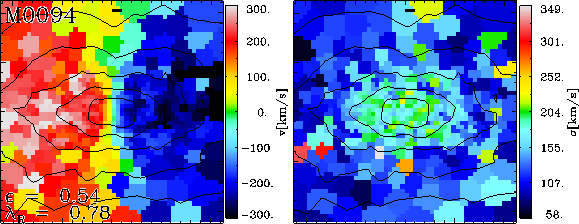}
\end{minipage} $\quad$
\begin{minipage}{0.47\textwidth}
\centering
retrograde $z$-tubes ($12.9\%$)\vspace{0.1cm}

\includegraphics[width = 0.99\textwidth]{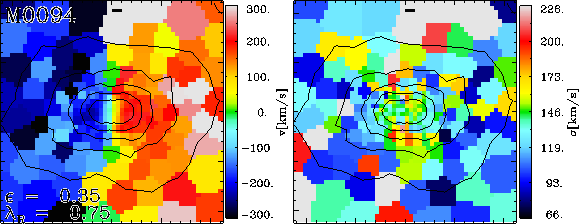}
\end{minipage}
\caption{The line-of-sight velocity and velocity dispersion maps of
  M0094 (galaxy class E) and its orbital components like for M0408 in
  Fig.~\ref{fig:LOSVD_decomp_M0408}. The galaxy has a
  counter-rotating core visible in the line-of-sight velocity (upper left 
  panel) which is generated by the $z$-tube population (middle left panel) 
  consisting of prograde and retrograde orbits (bottom panels). The 
  overall kinematics of the system is determined by the dominant 
  non-rotating box orbit population (upper right). This galaxy also has a 
  sizable $x$-tube component (12.9 \%).} 
\label{fig:LOSVD_decomp_M0094}
\end{figure*}

In Fig.~\ref{fig:orbits_LOSVD} we directly compare the orbital
structure of the six example galaxies to the two-dimensional (edge-on)
line-of-sight velocity and velocity dispersion maps up to $1.5
R_\mathrm{eff}$. $z$-tubes are the only orbits with a definite sense of 
rotation around the minor axis, and one expects a high line-of-sight
velocity (along the major-axis) at $z$-tube dominated radii. This can be 
seen for M0408 and M0163.

However, not all particles on $z$-tube orbits have to rotate in the same 
direction and the net-rotation can cancel out, resulting in slow rotation 
like for M0664, M0190 and M0175, with a slightly higher velocity 
dispersion. To investigate this effect, we separated the $z$-tubes into 
prograde (co-rotating) and retrograde (counter-rotating) orbits and plot 
the prograde $z$-tube fractions (green dashed lines in the orbit fraction 
plots) in Fig.~\ref{fig:orbits_LOSVD}. Wherever they dominate the 
$z$-tube population (M0408, and M0163), the galaxy is clearly rotating.
Similar trends also hold for $x$-tubes, however, M0190 is the only galaxy 
in the sample with noticeable major-axis rotation, which is why we 
concentrate on minor axis rotation.

We demonstrate the direct connection between the orbital composition
and the line-of-sight kinematics by performing a separate analysis for
stars on different orbit classes. In Fig.~\ref{fig:LOSVD_decomp_M0408}
we show the line-of-sight kinematics for M0408 for the full stellar
population, and for the stars on box orbits, $z$-tubes, and $x$-tubes 
respectively (from top left to bottom right). For this galaxy the rotation 
originates from the dominant population (66.4 \%) of $z$-tubes, whereas 
boxes (23.9 \%) and $x$-tubes (1.9 \%) show no (significant) rotation, 
typical for most fast rotators. Also the disk-like component, visible in the
isodensity contours, is composed of stars on $z$-tubes. The box orbits 
contribute to the central dispersion, whereas the dispersion at
$R_{\mathrm{eff}}$ is almost entirely determined by $z$-tubes.  
   
Slow rotators like M0175 (Fig.~\ref{fig:LOSVD_decomp_M0175}) have 
a non-rotating $z$-tube component (equally important co- and 
counter-rotating populations, see Fig.~\ref{fig:ztubes_rotation}) with a 
relatively high line-of-sight velocity dispersion (peaking at about 
280~km/s at $\sim 0.5 R_{\mathrm{eff}}$ with a drop in the center, see 
also Fig.~\ref{fig:ztubes_rotation}). Also the $z$-tubes show a 
characteristic box shape, i.e.\ they are not as flat as the one shown in 
Fig.~\ref{fig:classes_traj}, but rather `opened' in the edge-on projection 
(which is a typical behavior for tubes in non-axisymmetric potentials, 
see also \citealp{2005MNRAS.360.1185J}). Again, the nuclear dispersion is 
driven by the (very flattened) box orbit population. The system has a 
significant $x$-tube population (20.2\%) contributing to the dispersion 
around one $R_{\mathrm{eff}}$.

\begin{figure}
\centering
\includegraphics[width = 0.45\textwidth]{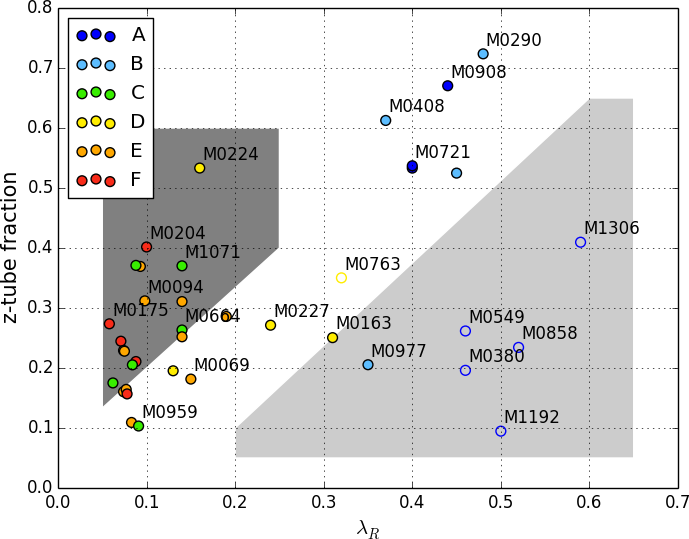}
\vspace{0.15cm}

\includegraphics[width = 0.45\textwidth]{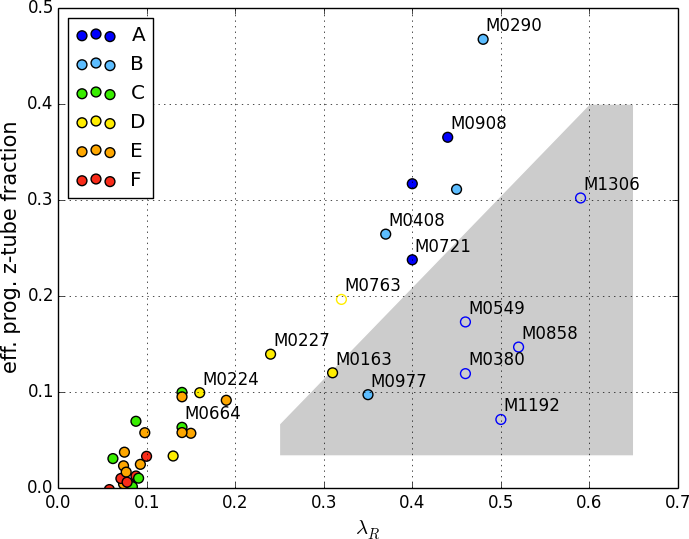}
\caption{{\it Upper panel:} The $z$-tube fraction (within
  $R_\mathrm{eff}$) of the simulated galaxies versus the
  $\lambda_R$-parameter. There is no correlation. Many low angular
  momentum galaxies have high $z$-tube fractions (dark grey shaded area)
  and high angular momentum systems can have low $z$-tube fractions
  (light grey shaded area). {\it Bottom panel:} The effective prograde
  $z$-tube fraction correlates nicely with $\lambda_R$ for most of the
  galaxies. The galaxies below the linear correlation (in the
  light-grey shaded triangle) are roughly those, for which we find
  signs of significant figure rotation (unfilled circles).}
\label{fig:ztubes_rotation}
\end{figure}

We have one simulated galaxy (M0094) with a counter-rotating core. 
For which we show the line-of-sight kinematics for the different orbit 
classes separated into prograde and retrograde $z$-tubes, boxes
and $x$-tubes. In the global line-of-sight velocity map (upper left
panel of Fig.~\ref{fig:LOSVD_decomp_M0094}) the counter-rotating core
($r \lesssim 0.3 R_\mathrm{eff} \simeq 2.2~\text{kpc}$) is visible
enclosed by the central isodensity contour. The left part of the nucleus 
has a negative velocity (light blue) of $\sim -50$~km/s and a positive
velocity at the opposite side. At larger radii the velocities change
sign. Again, the $z$-tubes alone carry a moderate amount of global 
rotation but the counter rotating core becomes clearly visible (middle 
left panel of Fig.~\ref{fig:LOSVD_decomp_M0094}). It is not a distinct
subsystem but generated by two extended counter-rotating $z$-tube
components. The center is dominated by retrograde $z$-tubes---the 
counter rotating core---larger radii ($r \gtrsim 0.5 R_\text{eff})$ become 
dominated by prograde $z$-tubes (see Fig.~\ref{fig:orbits_LOSVD}). 
Overall, however, the system is dominated by box orbits ($\sim 42\%$ 
of the stars are on box orbits). The characteristics of this simulated 
counter-rotating core resembles the only observed and modeled 
system with a counter-rotating core, NGC 4365 
\citep{Bosch_etal_2008}. This system, however, is dominated by tube 
orbits and the core is more clearly visible.

In the upper panel of Fig.~\ref{fig:ztubes_rotation} the (global)
$z$-tube fraction of the galaxies is plotted against their
$\lambda_R$-parameter. Surprisingly, there is almost no
correlation. In particular, galaxies with low $\lambda_R$ can have
$z$-tube fractions as high as 0.5. As mentioned before, the global
rotation of a galaxy can be low if the angular momentum on prograde
and retrograde $z$-tubes cancels out. This is the case for galaxies with
low $\lambda_R$ in the dark grey region in
Fig.~\ref{fig:ztubes_rotation}, mostly slow rotators of classes C, D, E and F.

A better measure for the amount of streaming motion is the `effective
prograde $z$-tube fraction', which we define as the normalized
difference of the  fraction of prograde and retrograde $z$-tubes: 
\begin{align}
\frac{(\text{\# prog. $z$-tubes}) - (\text{\# retrog. $z$-tubes})}{\text{\# all orbits}}. \label{eq:eff_pro_z}
\end{align}
Plotting this fraction against $\lambda_R$ (lower panel of
Fig.~\ref{fig:ztubes_rotation}) brings down the galaxies from the
dark grey area onto a tight correlation with $\lambda_R$ for most
galaxies. This indicates that for most simulated galaxies the $z$-tube orbit
family (and their separation into prograde and retrograde orbits)
determines the global rotation properties of the galaxies.

Galaxies, which do not follow the correlation of $\lambda_R$ and the 
effective prograde $z$-tube fraction (those in the light-grey area in 
Fig.~\ref{fig:ztubes_rotation}) are mostly of galaxy class A (fast-rotators 
with gas-rich minor mergers) and are also those, for which we found 
clear signs of figure rotation (indicated by open circles in 
Fig.~\ref{fig:ztubes_rotation}). Although the orbit classification for those 
galaxies is uncertain, we strongly suspect that the high $\lambda_R$ 
values have a significant contribution from figure rotation. The 
non-rotating galaxies for which we trust our classification, the effective 
prograde $z$-tube fraction correlates nicely with the 
$\lambda_R$-parameter.

\subsection{Orbits and LOSVD asymmetries}

\begin{figure}
\centering
\textbf{M0227}
\vspace{0.05cm}

\begin{minipage}{0.23\textwidth}
\includegraphics[width = 0.98\textwidth]{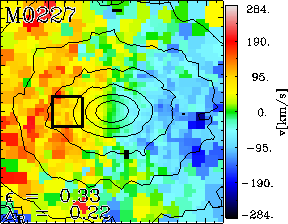}
\end{minipage}
\begin{minipage}{0.23\textwidth}
\includegraphics[width = 0.98\textwidth]{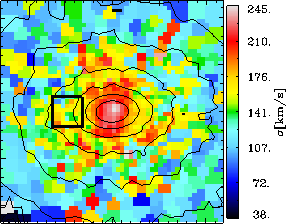}
\end{minipage}
\vspace{0.05cm}

\begin{minipage}{0.23\textwidth}
\includegraphics[width = 0.98\textwidth]{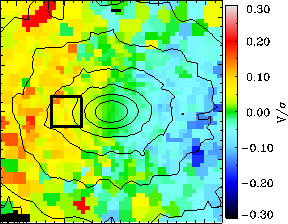}
\end{minipage}
\begin{minipage}{0.23\textwidth}
\includegraphics[width = 0.98\textwidth]{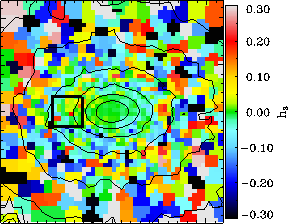}
\end{minipage}
\vspace{0.10cm}

\includegraphics[width = 0.46\textwidth]{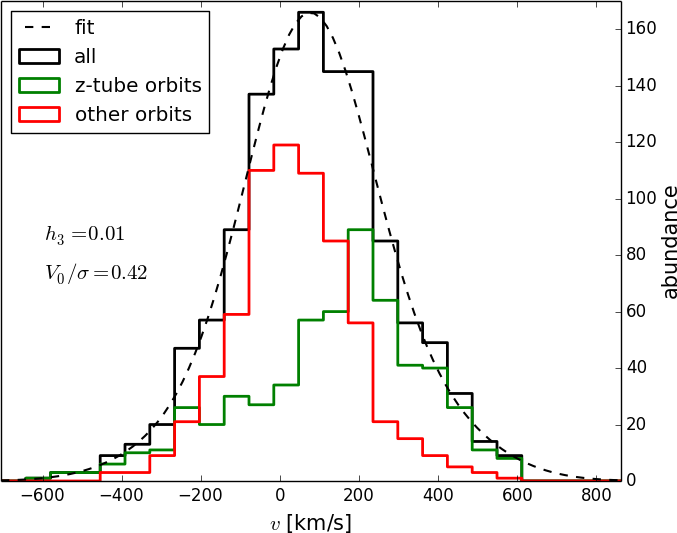}
\caption{Characteristic line-of-sight velocity distribution for all stellar 
  particles (bottom panel, black histogram) at the major axis around 
  $0.6 R_{\mathrm{eff}}$ ($0.4 \, R_\text{eff} < x < 0.8 \, R_\text{eff}$
  and $-0.2 \, R_\text{eff} < z < 0.2 \, R_\text{eff}$) for the fast
  rotating galaxy M0227 (class D). The investigated region is
  indicated in the line-of-sight velocity, dispersion, $V_0/\sigma$
  and $h_3$ maps (top panels) as black squares. The black dashed line
  indicates the Gauss-Hermite fit to the LOSVD with $h_3 = 0.01$. The 
  LOSVD of stars on $z$-tube orbits (green
  histogram) peaks at $v \approx 200~\text{km/s}$, but the total 
  LOSVD is dominated by particles which are mostly on box orbits (red  
  histogram) peaking at $v \approx 0$. Such a profile is typical for fast 
  rotating galaxies formed in dissipationless mergers
  \citep{2001ApJ...555L..91N, 2006MNRAS.372..839N, 
  Naab_etal_2013}.  
}
\label{fig:LOSVD_eff_M0227}
\end{figure}

\begin{figure}
\centering
\textbf{M0408}
\vspace{0.05cm}

\begin{minipage}{0.23\textwidth}
\includegraphics[width = 0.98\textwidth]{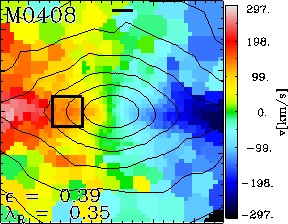}
\end{minipage}
\begin{minipage}{0.23\textwidth}
\includegraphics[width = 0.98\textwidth]{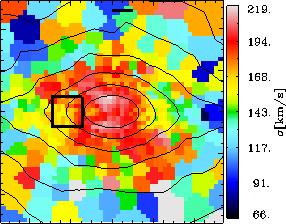}
\end{minipage}
\vspace{0.05cm}

\begin{minipage}{0.23\textwidth}
\includegraphics[width = 0.98\textwidth]{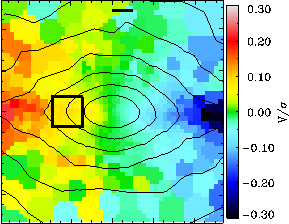}
\end{minipage}
\begin{minipage}{0.23\textwidth}
\includegraphics[width = 0.98\textwidth]{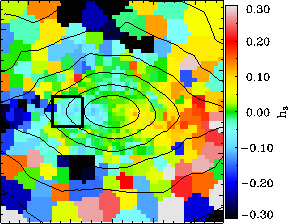}
\end{minipage}
\vspace{0.10cm}

\includegraphics[width = 0.46\textwidth]{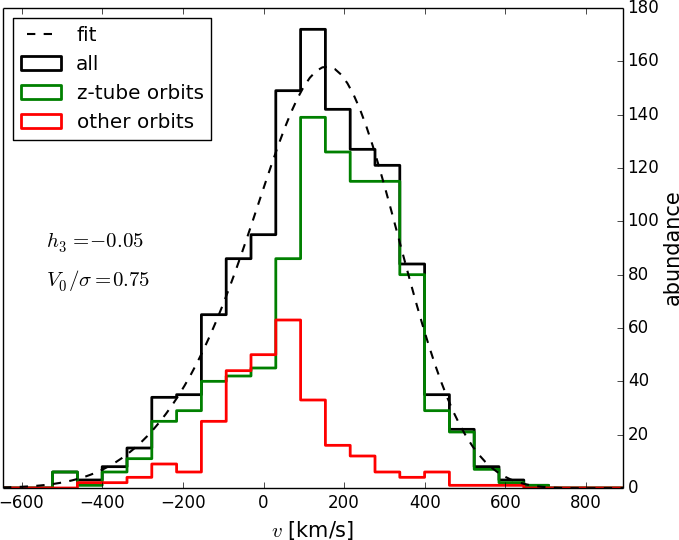}
\caption{Same as Fig.~\ref{fig:LOSVD_eff_M0227} but for the fast
  rotating galaxy M0408 (Class B). Here the stellar orbits in the observed
  region are dominated by tubes (green histogram) generating an overall
  LOSVD with a steep leading wing (black histogram and dashed line,
  $h_3 = -0.05$). Other orbit classes (red histogram) are
  subdominant. Typically high tube-to-box fractions generate LOSVDs
  with steep leading wings
  \citep{2006MNRAS.372..839N,2010ApJ...723..818H} and such a LOSVD is
  typical for fast rotating galaxies that have formed under the
  significant influence  of dissipation (see e.g. class A and B
  in \citealp{Naab_etal_2013}).}   
\label{fig:LOSVD_eff_M0408}
\end{figure}

For most observed fast rotating galaxies the amplitude of the 
third-order component ($h_3$) of a Gauss-Hermite fit to the 
line-of-sight velocity profile (Eq.~\ref{eq:gauss_hermite}) anti-correlates 
with rotational support ($V_0 / \sigma$; \citealp{Bender_etal_1994, 
Halliday_etal_2001, Pinkney_etal_2003, 2013MNRAS.432.1768K}; for 
the $V_0 / \sigma$ data from the ATLAS$^\text{3D}$ sample see 
\citealp{2011MNRAS.414.2923K, Krajnovic_etal_2008}). The velocity 
profile has a steep leading and a long trailing wing. This holds for oblate 
systems resembling tow-integral models \citep{1994MNRAS.268.1019D, 
Bender_etal_1994} and/or can be indicative of an embedded disk 
component \citep{van_der_Marel_Franx_1993, Bender_etal_1994, 
Fisher_1997,2001ApJ...555L..91N, 2013MNRAS.432.1768K}. In this 
section we use the cosmological galaxy formation simulations to explain 
the origin of the anti-correlation of $h_3$ and $V_0$ and its connection 
to the orbital structure.  

\citet{Naab_etal_2013} found that only fast rotating galaxies with late 
gas-rich mergers or late dissipation (classes A and B in their 
classification) show a $h_3$-$V_0/\sigma$ anti-correlation. These are 
also the galaxies with the highest $\lambda_R$-parameter and, as we 
have shown here, those with a the highest effective prograde $z$-tube 
fraction. 

It has been shown with idealized experiments that gas dissipation
(cooling gas flow to the center, triggered by a merger event) can
explain this behavior. In collisionless mergers the stellar
population of the remnant is dominated by stars on box orbits and even
if the remnant is rotating the LOSVD has steep trailing wing (positive
$h_3$) which is hardly observed \citep{2001ApJ...555L..91N}. Gas
inflow during a merger, however, leads to centrally concentrated, 
axisymmetric potentials, suppressing the population of box orbits
\citep{1996ApJ...471..115B}. With now more stars on $z$-tubes (which 
carry the angular momentum) the LOSVD gets a steep leading wing 
and the observed anti-correlation of $h_3$ and $V_0/\sigma$ can be 
recovered \citep{2006MNRAS.372..839N}. This is supported by the 
formation of new stars in a re-forming disk 
(\citealp{2006MNRAS.372..839N, 2009ApJ...705..920H, 
2010ApJ...723..818H} and also see \citealp{Bender_etal_1994}).

Here we present the same effect for two galaxies formed in a 
cosmological context. In the top panels of
Fig.~\ref{fig:LOSVD_eff_M0227}) we show the two-dimensional maps of
$V_0, \sigma, V_0/\sigma$, and $h_3$ for M0277, a fast rotating galaxy
that has experienced a late gas-poor (collisionless) major merger
(typical for the class D galaxies in \citealp{Naab_etal_2013}). For
this galaxy the LOS velocity and $h_3$ are correlated. In the bottom 
panel of Fig.~\ref{fig:LOSVD_eff_M0227} we show, as an example, the 
LOSVD at $\sim \! 0.6 R_{\mathrm{eff}}$. It is dominated by non-rotating 
stars ($73\%$) without net-rotation---mostly on box orbits. The stars on 
prograde $z$-tubes shift the peak towards positive velocities. The resulting 
distribution is almost symmetric as not only the prograde $z$-tubes (not as
dominant as the box orbits) broaden the distribution towards positive 
values, but there are also some retrograde $z$-tubes, that do the same 
towards negative velocities. Hence, only a slightly positive value for 
$h_3$.

For M0408 the situation is different. This galaxy is a fast rotator with 
late gas-rich major merger and it shows a clear anti-correlation of the 
LOS velocity and $h_3$ (top panels Fig.~\ref{fig:LOSVD_eff_M0408}). 
A typical LOSVD is shown in the bottom panel of 
Fig.~\ref{fig:LOSVD_eff_M0408}. At the same radius this galaxy is 
dominated by high angular momentum $z$-tube orbits which by
themselves already generate a LOSVD with a steep leading wing. The
broad trailing wing hosts stars on retrograde tubes. Stars on other
orbits without angular momentum do not affect the shape of the LOSVD
very much.

If we restrict ourselves to galaxies without figure rotation, we can 
conclude that all orbits apart from $z$-tubes have no intrinsic 
angular momentum (around the minor axis) and hence their LOS 
velocity profile is almost symmetric and peaks around $v = 0$. The 
$z$-tube orbits can be approximated as two peaks: one from retrograde 
and one from prograde $z$-tubes. For a rotating system (around the minor 
axis) the latter peak has a larger amplitude. If the other orbits (centering 
at $v= 0$) are subdominant (like in Fig.~\ref{fig:LOSVD_eff_M0408}) the 
overall distribution then peaks with the prograde $z$-tubes has a trailing 
wing that is broadened by the retrograde $z$-tubes and other orbits. If the 
other orbits, however, are dominant (like in Fig.~
\ref{fig:LOSVD_eff_M0227}), the overall distribution peaks in-between 
the prograde $z$-tubes and the other orbits and is rather symmetric.

\subsection{Signatures of two phase assembly}

All galaxies presented here consist of two populations of stars. One has
formed within the galaxy (the {\it in situ} component) and the second has
formed in other galaxies and have been accreted in mergers (the 
accreted component). In general, {\it in situ} formation dominates at high
redshift and the accretion of stars becomes more important at low
redshift, and more so for massive systems
\citep{2008MNRAS.384....2G, Oser_etal_2010, 2012MNRAS.425..641L, 2013MNRAS.428.3121M, 2013IAUS..295..340N}. The separate origin 
of these components might also results in different orbit populations. 

In Fig.~\ref{fig:insitu_classes} we plot the orbit fractions of the {\it in situ} 
and accreted component as well as for the dark matter particles as a 
function of radius for four characteristic examples. Surprisingly, most 
galaxies show little difference in the orbital composition of these 
components (M0163 and M0175). Possible small difference might be washed out, 
since for individual particles the orbit classification is uncertain on the 
per-cent level (see Sec.~\ref{orbit_stability}). For M0664 and M0917, 
however, the orbit composition of the {\it in situ} component is clearly distinct.

The former has a tube biased {\it in situ} component and has a recent 
gas-poor major merger, where the different dynamical history of the two 
merged galaxies is still imprinted in the orbit structure. The latter, 
M0977, had a late gas-rich major merger, in which new stars formed 
the dissipative component, and and thus it has a prominent late {\it in situ} 
formed component that is populating mostly $z$-tubes (except for the 
inner radii, $r \lesssim 0.5 R_\text{eff}$). The orbital structure of the 
dark matter component, however, is always very similar to the accreted 
component, independent of the different assembly histories of the 
galaxies. It is plausible to assume that frequent mergers, which are 
relevant for galaxy classes C, D, E, \& F, mix the {\it in situ} and accreted 
stellar populations efficiently and the dark matter particles behave in a
similar way.

In addition the orbit analysis might not be the best diagnostics. Although 
$z$-tubes are all centrophobic they can significantly differ in shape from 
circular to very eccentric. A good measure for this behavior is provided 
by the anisotropy $\beta$ \citep{Binney_Tremaine_2nd}:
\begin{align}
\beta \equiv 1 - \frac{\langle v_\phi^2 \rangle + \langle v_\theta^2 \rangle}{2 \, \langle v_r^2 \rangle},
\end{align}
where 
$v_r$, $v_\phi$, and $v_\theta$ are the velocities in spherical 
coordinates. This parameter is zero for isotropic motion, positive if the 
velocities are radial biased and negative if they are tangentially biased.

We plot the corresponding radial anisotropy profiles of M0163, M0175, 
M0664, and M0977 in the right panels of Fig.~\ref{fig:insitu_classes} and 
also separate the {\it in situ} component and accreted component as well as 
the dark matter particles. For M0664 and M0977 the orbital structure is 
reflected in the anisotropy profiles: where in galaxy M0664 the {\it in situ} 
stars populate box orbits more than the accreted stars do, there the 
motion of the {\it in situ} stars is also more radially biased than of the 
accreted stars; similarly for M0408, where the {\it in situ} stars are more 
tangentially biased, they preferentially populate $z$-tubes. However, we 
also see different anisotropy profiles for {\it in situ} and accreted stars in 
galaxies which show no significant difference in the orbit profiles, e.g.\ 
for M0175. Here the accreted stars are moderately radially biased 
($\beta \approx 0.4$) whereas the {\it in situ} stars---having almost identical 
orbital structure---are isotropic. But this is not true for all galaxies. There 
are some that have alike $\beta$-profiles for {\it in situ} and accreted stars 
(e.g.\ M0163).

We also investigated the dark matter anisotropy profiles (black lines in 
Fig.~\ref{fig:insitu_classes}) and they turned out to be very isotropic or 
sometimes mildly radially biased (cf.\ M0977) where they have a 
tendency to follow the accreted component. To see whether this is 
universal and to identify trends in the anisotropies, we plotted the 
$\beta$-profiles of the different galaxy classes (except for those galaxies 
that show a massive near-by substructure that would compromise the 
results) in Fig.~\ref{fig:beta_gal_classes}. The dark matter particles are 
indeed always isotropic to mildly radially biased, with no trend along the
different galaxy classes. The profiles of the other components, however, 
have clear dependencies on the assembly history of the galaxies.

\begin{figure*}
\centering
\raisebox{-.5\height}{
\includegraphics[width = 0.345\textwidth]{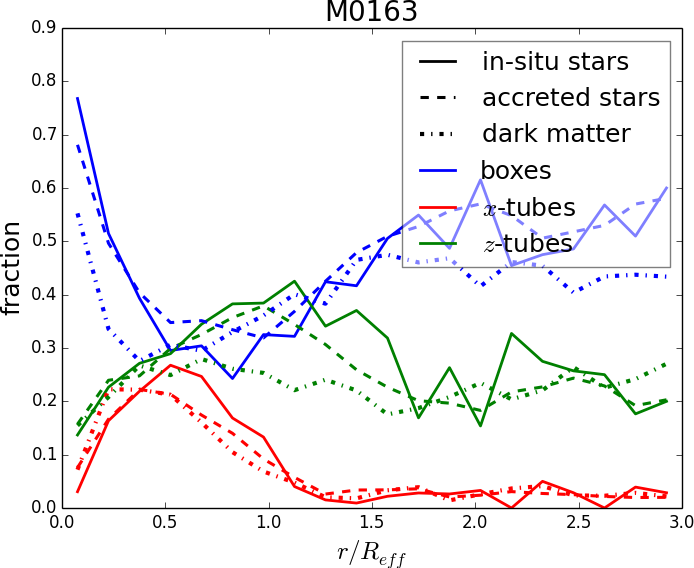}
$\quad$
\includegraphics[width = 0.355\textwidth]{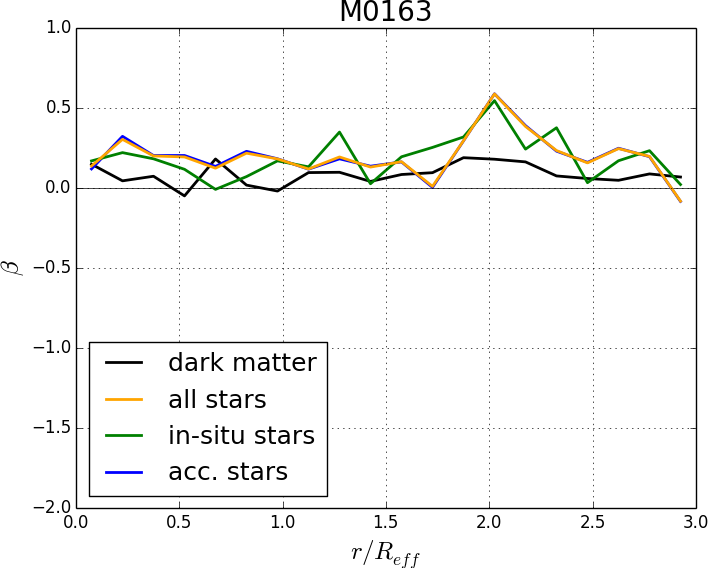}
} $\quad$ Class D
\vspace{0.15cm}

\raisebox{-.5\height}{
\includegraphics[width = 0.345\textwidth]{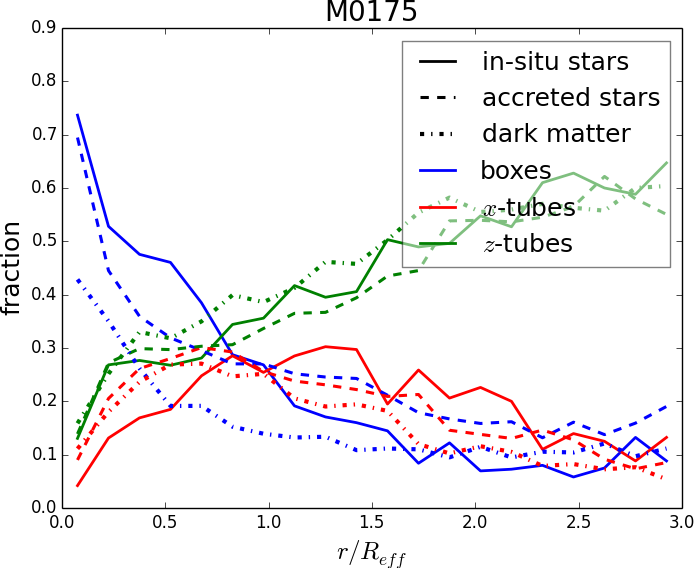}
$\quad$
\includegraphics[width = 0.355\textwidth]{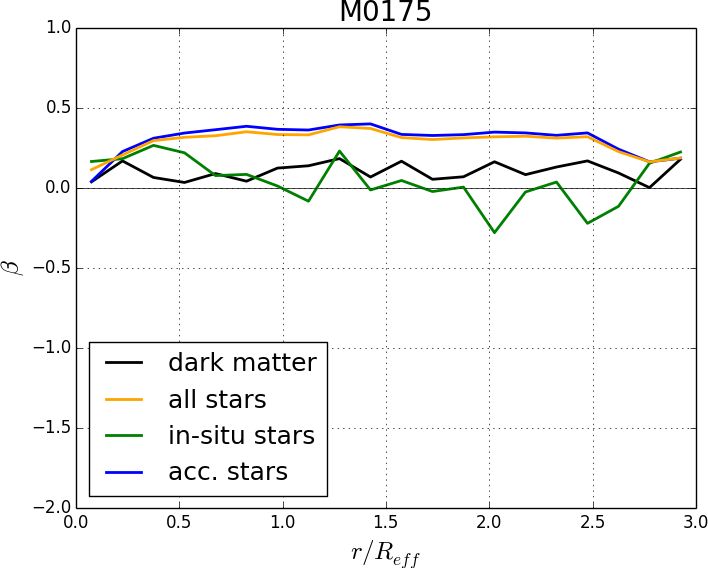}
} $\quad$ Class F
\vspace{0.15cm}

\raisebox{-.5\height}{
\includegraphics[width = 0.345\textwidth]{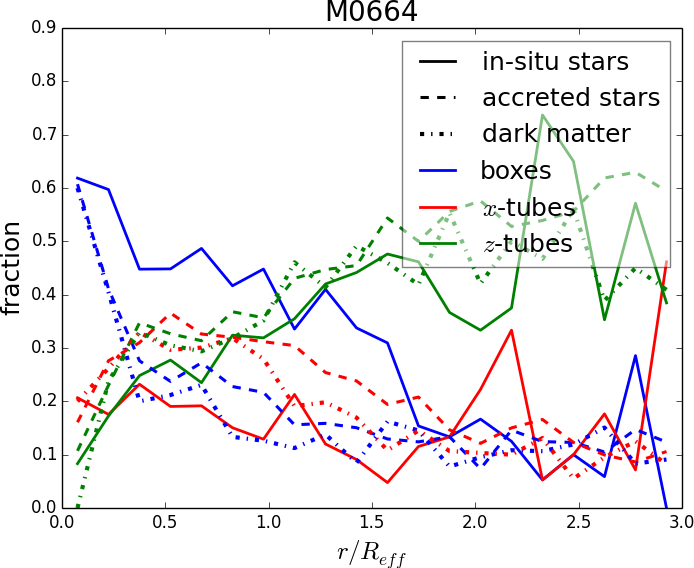}
$\quad$
\includegraphics[width = 0.355\textwidth]{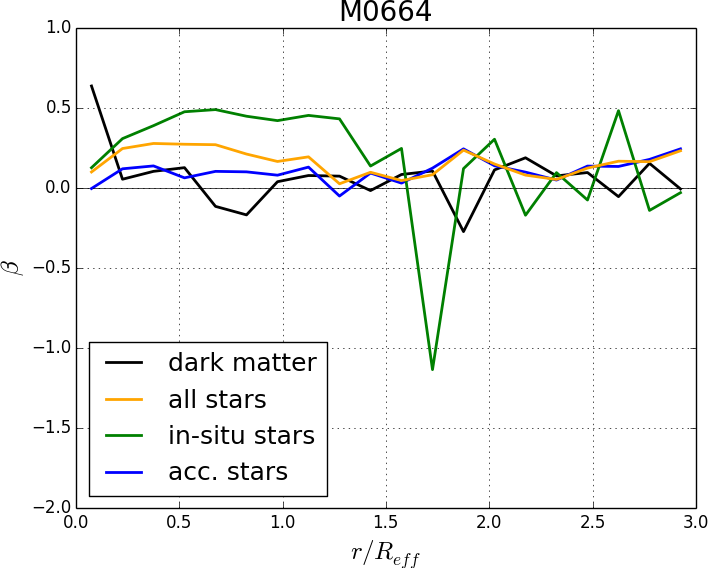}
} $\quad$ Class E
\vspace{0.15cm}

\raisebox{-.5\height}{
\includegraphics[width = 0.345\textwidth]{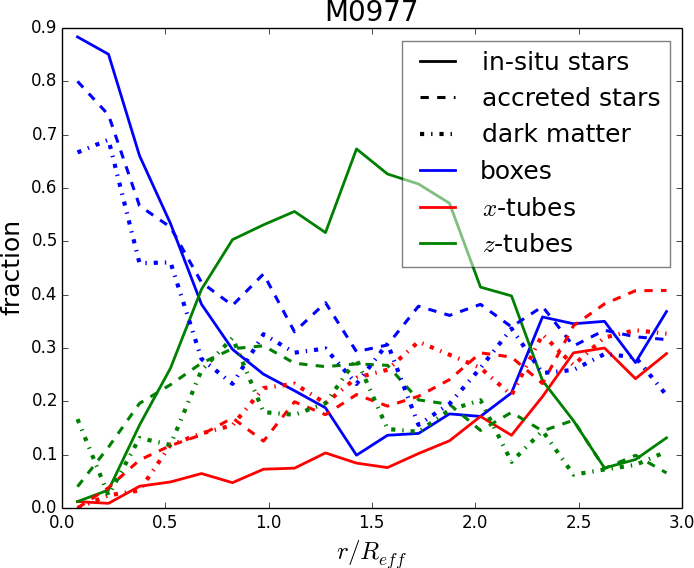}
$\quad$
\includegraphics[width = 0.355\textwidth]{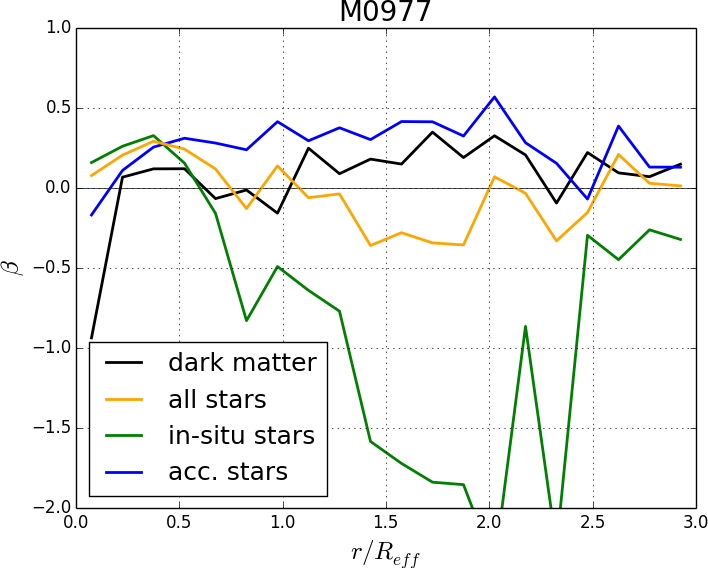}
} $\quad$ Class B
\caption{{\it Left panels}: The radial orbital structure of the
  {\it in situ} stars (solid lines) and the accreted stars (dashed lines) as
  well as dark matter particles (dash-dotted lines) for four galaxies
  (M0163, M0175, M0664 , and M0977). For most galaxies with
  collisionless recent assembly histories (classes D, E, and F) there
  is almost no difference between the {\it in situ} and accreted component
  and the dark matter particles. If there are mild variations (like
  for the box population in M0664) the accreted stars and dark matter
  are more alike. This is also true for M0977, but this galaxy has
  relevant late {\it in situ} star formation (and no late large mergers)
  forming a prominent $z$-tube population dominating most of the
  galaxy. {\it Right panels:} The anisotropy parameter $\beta$ as a
  function of radius for dark matter (black), all stars (yellow), as
  well as separated into the {\it in situ} (green) and accreted (blue)
  component. The dark matter is in general isotropic at all
  radii. M0163 is dominated by box orbits at all large radii and
  radially biased ($\beta \sim 0.2$). M0175 is dominated by $z$-tubes at
  large radii but here the {\it in situ} stars are more isotropic that the
  accreted stars which are also on $z$-tubes but clearly radially
  biased. For M0977 this effect is even stronger with a clearly
  tangentially biased {\it in situ} component.}
\label{fig:insitu_classes}
\end{figure*}

\begin{figure*}
\centering
\includegraphics[width = 0.32\textwidth]{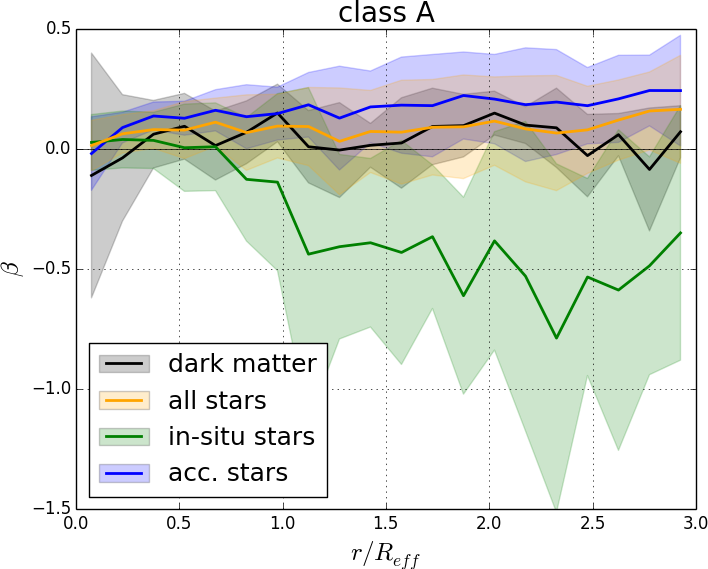} \ 
\includegraphics[width = 0.32\textwidth]{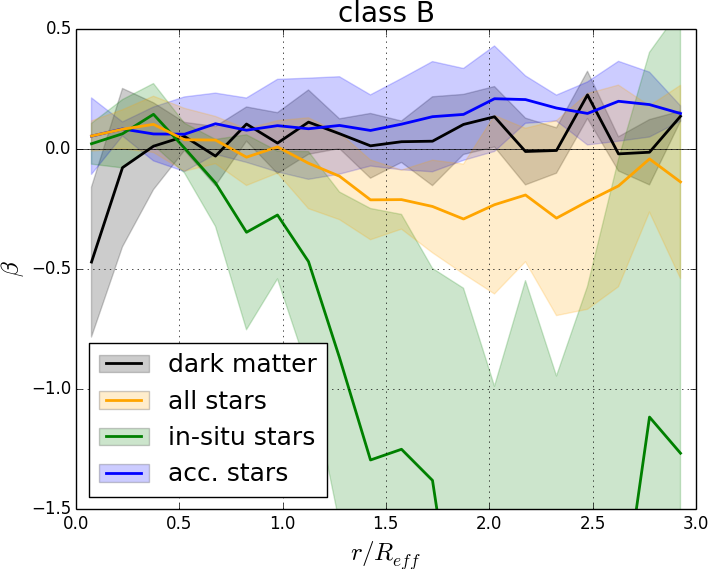} \ 
\includegraphics[width = 0.32\textwidth]{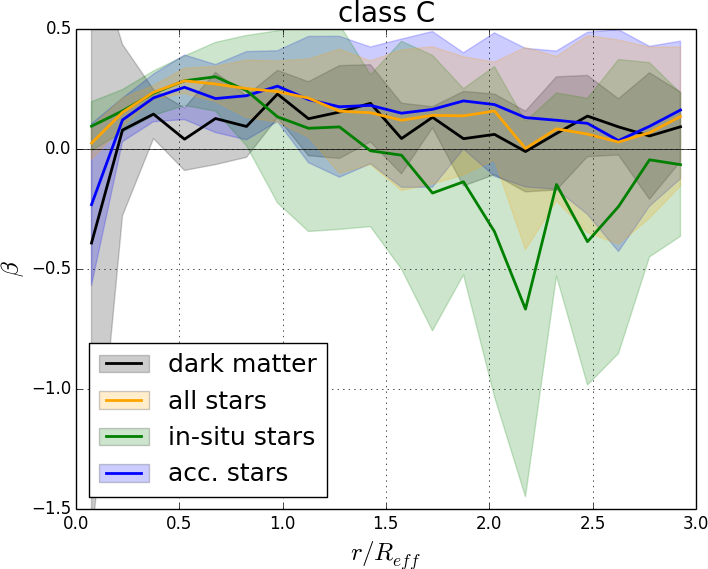}
\vspace{0.15cm}

\includegraphics[width = 0.32\textwidth]{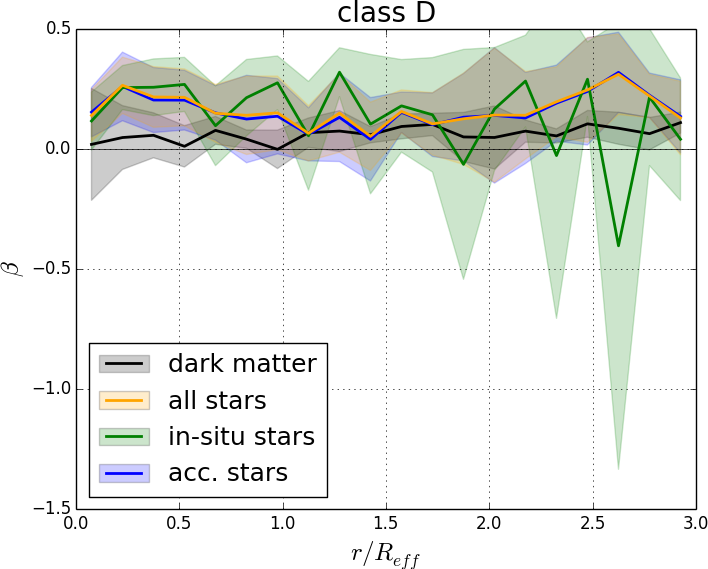} \ 
\includegraphics[width = 0.32\textwidth]{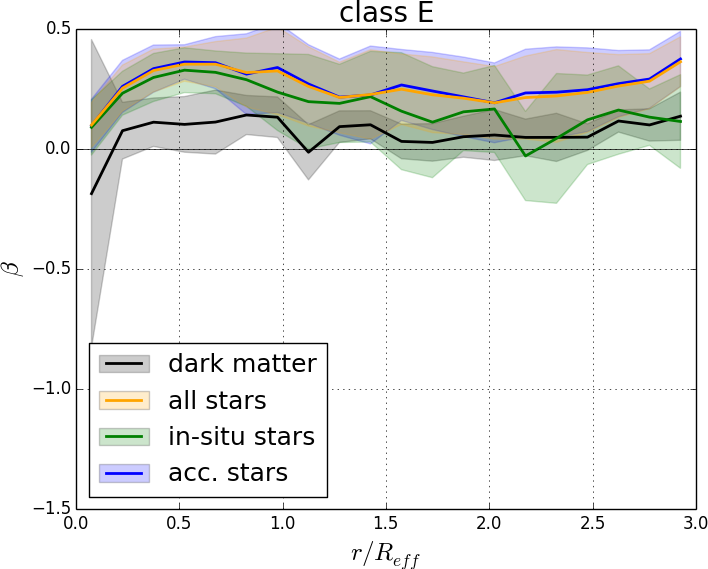} \ 
\includegraphics[width = 0.32\textwidth]{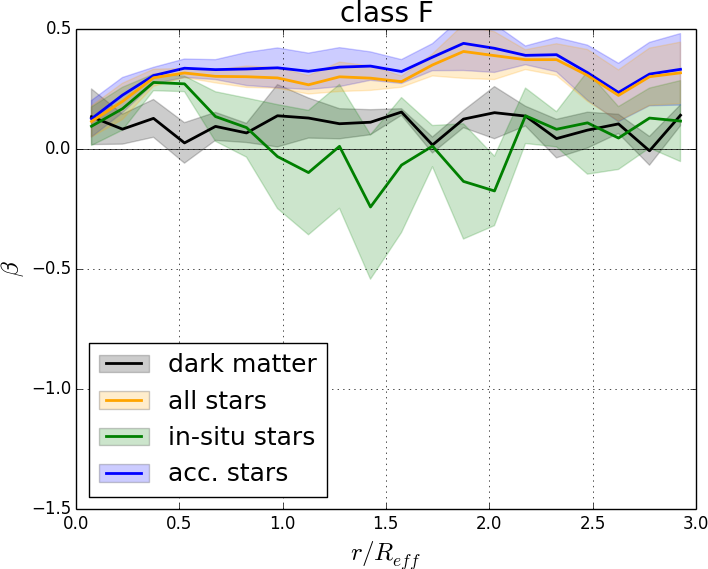}
\caption{The mean anisotropy parameter profile (line) for dark matter
  and stars as well for the {\it in situ} and accreted components only is 
  plotted for the different galaxy classes. Three galaxies---M0069 (E), 
  M0162 (E) and M0948 (F)---that have heavy nearby substructures 
  (indicating an on-going/up-coming major merger) are excluded from 
  these calculations. The pale bands around the mean indicate the 
  standard deviation among the galaxy classes.
}
\label{fig:beta_gal_classes}
\end{figure*}

We find the {\it in situ} component to have smaller $\beta$ than the accreted 
component in general and in agreement with
\cite{2014MNRAS.438.2701W}, who 
grouped the galaxies among their {\it in situ} fractions and plotted the overall 
stellar anisotropy profiles for all galaxies. This is not surprising, since 
{\it in situ} stars form from the dissipative gas component, that settles down 
onto a rotating disk conserving angular momentum. The rotating 
dynamical element is also imprinted in the {\it in situ} stars that then should 
preferentially populate $z$-tubes---the only ones with intrinsic rotation. The 
accreted star component has fallen in from all directions and, hence, are 
expected to be on more radially biased orbits.

We furthermore find that fast-rotators with late gas-rich mergers (classes 
A and B) have very tangentially anisotropic {\it in situ} components, 
especially beyond the effective radius. Accreted stars, however, are a bit 
radially biased and hence the overall stellar anisotropy is slightly radially 
biased for class A galaxies---those with late gas-rich minor 
mergers---anyway. Class~B galaxies---those with late gas-rich major 
mergers---have similar anisotropies of the accreted stars, the {\it in situ} 
component, however, is even stronger tangentially biased than it is for 
class~A galaxies. The importance of the {\it in situ} stars yields a tangentially 
biased stellar anisotropy.

Interestingly, the {\it in situ} stars are always slightly radially biased or 
isotropic at most for fast-rotators with late gas-poor major mergers 
(class D). Slow-rotators with gas-rich major mergers (class C) have very 
similar stellar anisotropy profiles. The {\it in situ} star, however, are mildly 
tangentially biased at large radii in contrast to those galaxies with 
gas-poor major mergers (classes D and E). We, hence, see that the gas 
fraction plays an about equally important role in determining the 
anisotropy of the in situ stars and we consequently find the most radially 
biased motion of the stars in slow-rotators with gas-poor major/minor 
mergers (classes E and F). For those with major mergers (class E) even 
the {\it in situ} stars are no longer tangentially biased and they are roughly 
isotropic for those with minor mergers.

It seems as the anisotropy parameter $\beta$ is better correlated with 
the {\it in situ} fractions and the quantized nature of orbit classification can 
often not capture the different assembly histories of {\it in situ} and accreted 
stars. Moreover, $\beta$ is better observable, and although the most 
prominent trends are at large radii ($r \gtrsim R_\text{eff}$), the 
differentiation between fast-rotators with gas-rich merger histories 
(classes A and B) and the other galaxy classes is already strong at 
smaller radii (compare yellow lines for stellar $\beta$ in 
Fig.~\ref{fig:beta_gal_classes}).

We see that the anisotropy indeed reveals more about the assembly 
history of the galaxies than the orbit classes do, but they also 
complement one another as the rotation of the galaxies is not directly 
reflected in the anisotropy profiles. The dark matter particles have very 
similar orbital structures as the stellar component (see 
Fig.~\ref{fig:insitu_classes}), but they are very isotropic for all galaxies 
among our ensemble as it can be seen from 
Fig.~\ref{fig:beta_gal_classes}. The anisotropies we find are also consistent with observational results:
there is a relatively large variety of anisotropies \citep{2007MNRAS.382..657T} but mostly mildly radially biased ones \citep{2008MNRAS.385.1729D, 2011MNRAS.415.1244D} the only strong exception is fast-rotating galaxies with late gas-rich major mergers (class B) beyond the effective radius.

We plot the global fraction of stars that formed {\it in situ} since $z \approx 
2$ against the anisotropy $\beta$ in Fig.~\ref{fig:aniso_insitu}. {\it In situ} 
stars that have formed earlier have undergone multiple merger events 
and, hence, the fingerprint of their dynamical history in the anisotropy is 
probably washed out. We indeed see a weak correlation, though only for 
the inner parts ($r \lesssim 0.5 R_\text{eff}$) of the galaxies 
(Fig.~\ref{fig:aniso_insitu}). For the region $0.5 \, R_\text{eff} < r 
< 1.0 \, R_\text{eff}$ there is already very little correlation left. This is 
mainly due to the fact that the {\it in situ} fractions in this region has 
decreased by a factor of almost two from that within $0.5 R_\text{eff}$, 
and hence they do not influence $\beta$ as much.

\begin{figure}
\centering
\includegraphics[width = 0.45\textwidth]{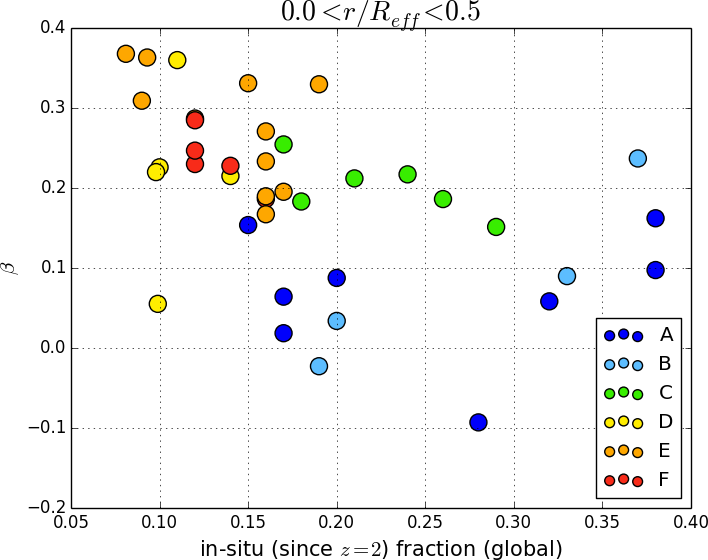}
\vspace{0.15cm}

\caption{For the inner part of the galaxies ($r < 0.5 R_\text{eff}$, upper panel) the anisotropy parameter $\beta$ correlates with the {\it in situ} fraction.
Furthermore, the galaxy classes from \protect\cite{Naab_etal_2013} separate in the diagram.
For larger radii, however, the correlation is nearly completely gone (see lower panel).}
\label{fig:aniso_insitu}
\end{figure}

Moreover, we see that the galaxy classes defined by \cite{Naab_etal_2013} separate in the anisotropy--{\it in situ} fraction diagram.
Fast-rotators with late gas-rich mergers or late dissipation (classes A and B) have rather high {\it in situ} fraction and are isotropic of slightly tangentially biased.
Slow-rotators with late gas-rich major mergers (class C) still have high {\it in situ} fractions due to the shocking gas in the mergers, but are more radially biased, since they do not rotate as much.
The remaining classes of fast-rotators with late gas-poor major mergers (class D) and slow-rotators with gas-poor mergers (classes E and F) are located in the upper left corner in the diagram with small {\it in situ} fractions and are tangentially biased.

\section{Summary and Discussion}
\label{summary}

We have presented an orbit analysis of star and dark matter particles
out to three effective radii for a sample of 42 galaxies formed in
cosmological zoom simulations of \cite{Oser_etal_2010}. With an
improved version of the \cite{Carpintero_Aguilar_1998}
spectral classification scheme we classified orbits of stars and dark
collisionless dark matter particles. For the stellar orbits we found
that box orbits and $z$-tubes are most abundant among the simulated
galaxies. The distribution of orbits with radius, however, can vary
significantly from galaxy to galaxy. A common feature is the
relatively high central ($r \lesssim 0.4 \, R_\text{eff}$) box orbit
fraction, much higher than what is found when triaxial Schwarzschild
models are applied to observed LOSV maps (c.f.\ NGC 4365,
\citealp{Bosch_etal_2008} and NGC 3379 and NGC
821,\citealp{2009MNRAS.398..561W}). It is likely that underestimating
the influence of dissipation of the assumed model  
(see the discussion in \citealp{Oser_etal_2010}) and/or neglecting
black hole impact play a significant role. Both (not included) effects will 
help to make the central region more axisymmetric and therefore
suppress the population of box orbits.  
 
Galaxies with figure rotation (6 out of 42), which have uncertain
orbit fractions, were identified and we excluded them from any
interpretation that uses orbit analysis. For the remaining galaxies we
show that LOS rotation, quantified by the $\lambda_R$-parameter,
originates from  streaming motions of stars. We demonstrate that the
value of $\lambda_R$  directly correlates with the `effective prograde
$z$-tube  fraction' (i.e.\ the difference of the fraction of prograde
$z$-tubes and the fraction of retrograde $z$-tubes normalized by the total
number of $z$-tubes). 

We find the expected correlations of box orbit, $z$-tube, and $x$-tube
fractions with galaxy triaxiality (e.g. \citealp{2005MNRAS.360.1185J}). 
$z$-tubes live in oblate systems, $x$-tubes in prolate ones and box orbits
are most abundant in very triaxial systems ($T \approx 0.5$). We
construct mock LOSV maps---also from single orbit families---of the
simulated galaxies and recover the observational impact of the
different orbit families. For example, rotation originates from
streaming motion of stars on $z$-tubes as expected from the
investigation of the correlation of the effective prograde $z$-tube
fraction and the $\lambda_R$ parameter. One galaxy has a
counter-rotating core and, similar to NGC 4365, we are able to
demonstrate that the core is not a kinematically decoupled system, but
originates from a superposition of the smooth distributions of
prograde and retrograde $z$-tubes (c.f.\ \citealp{Bosch_etal_2008}). 

Using the orbit classification, we also can also explain the origin of
the observed anti-correlation between $h_3$ and
$V_0/\sigma$. Many simulated fast rotators show and anti-correlation
but correlated $h_3$ and $V_0/\sigma$ is also possible. The group of
galaxies with enhanced dissipation, i.e. with more 'late' {\it in situ} star
formation due to gas accretion or gas-rich mergers, shows a clear
tendency for an anti-correlation \citep{Naab_etal_2013}. The increased
relative fraction of prograde tube orbits creates a steep leading wing in the 
LOSVD. Physically, this originates from a suppression of box orbits in
more axisymmeric potentials. Rotating systems with little late
dissipation do not show this anti-correlation. This conclusion is in
agreement with studies of isolated mergers 
\citep{2006MNRAS.372..839N,2009ApJ...705..920H,2010ApJ...723..818H}.

Using the six galaxy classes with different formation histories
defined in \cite{Naab_etal_2013} we only see weak trends of orbit
fractions with the formation history apart from systems with late
dissipation showing high $z$-tube fractions. Surprisingly, the
distribution of orbits of the {\it in situ} formed stars, accreted stars,
and dark matter particles are very similar, with the exception of
galaxies with a lot of late dissipation. This result compares well
with \cite{Bryan_etal_2012}, who find similar distributions for stars
and dark matter particles at larger radii. Only two of our galaxies
exhibit a clear distinction between the orbit profiles of {\it in situ}
formed and accreted stars. One system with a late gas-rich major
merger (M0977) has a higher fraction of $z$-tubes for the {\it in situ} formed
stars than for the accreted stars. This is expected as the {\it in situ}
stars form from the dissipative gas component which can settle onto an
axisymmetric, oblate disk before star formation. This is `memorized'
in the kinematic features and preferentially populates stars on $z$-tubes. 

We found that the velocity anisotropy (measured by the $\beta$
parameter) depends more on the formation history than orbit distribution.  
Galaxies with gas-rich mergers and gradual dissipation have can have 
mildly radially biased motions of the stars ($\beta \simeq 0.2$) and
tangentially biased motions of the stars (up to $\beta \simeq -0.5$),
especially at larger radii ($r \gtrsim R_\text{eff}$). Galaxies with
dry mergers always have radially biased motions of stars ($0.1
\lesssim \beta \lesssim 0.4$). Stars that form {\it in situ} from the
dissipative gas component tend to have tangentially biased motions,
whereas accreted stars fall in on radially biased trajectories and
hence preferentially end up on tangentially biased orbits.  We find
that {\it in situ} formed stars are mildly radially anisotropic ($\beta
\simeq 0.25$) for gas-poor major mergers remnants and otherwise
isotropic to strong tangentially anisotropic ($\beta < -1.0$) for
fast-rotators with gas-rich mergers. The dark matter particles,
however, are always close to isotropic. Orbit families are less well
correlated as for example tube orbits can become radially biased when
they become eccentric.

It is plausible to assume that {\it in situ} stars and accreted stars can have 
different metallicities but the accreted galaxies have lower mass (e.g.\ 
\citealp{1982MNRAS.199..493V, 2013MNRAS.429.2924H}). Therefore 
it will be interesting to investigate a connection between metallicities 
and kinematic properties (orbit families, anisotropy, etc.). 
Zoom-simulations reproducing the observed evolution of the 
mass--metallicity relation (e.g. \citealp{2013MNRAS.436.2929H, 
Aumer_etal_2013}) would be well suited for this. As our 
simulations do not follow metal evolution such an investigation is, 
however, beyond the scope of this paper.
However, orbit families do provide important information about the
formation processes. Both AGN feedback and stellar feedback can lead 
to a more or less dissipative formation of galaxies (see
e.g. \citealp{1998MNRAS.295..319M, 2012MNRAS.419.3200H, 
2013MNRAS.433.3297D, 2014arXiv1403.6124U}). 
Enhanced dissipation results in more axisymmetric and oblate galaxies
and with have higher $z$-tube fractions (the dissipative formation of
disk galaxies is an extreme example,
e.g. \citealp{Aumer_etal_2013}). For large radii
\cite{Bryan_etal_2012} could indeed show that feedback rises the
amount of $z$-tubes and reduced the amount of box orbits for star
particles as well as dark matter particles.  However, the impact of
dissipation might be particularly important for galaxy
centers. Furthermore, feedback can change merger histories (in general
reduces the number of minor mergers) and increases gas accretion and 
the amount of gas involved in mergers (e.g. 
\citealp{2012MNRAS.419.3200H, 2014arXiv1403.6124U}). Therefore
`feedback physics' might leave a clear fingerprint in the  orbital
composition of massive galaxies.

\section{Acknowledgements}

We thank Davor Krajnovi\'c for his comments and suggestions and 
thank the anonymous referee for valuable suggestions. Thorsten Naab 
acknowledges support by the DFG cluster of excellence `Origin and 
Structure of the Universe'.

\clearpage

\bibliographystyle{mn2e}
\bibliography{./references}


\label{lastpage}
\end{document}